\documentclass[11pt, a4paper]{article}
\usepackage{CJKutf8}
\usepackage{geometry}
\usepackage{subfigure}

\usepackage{amssymb}
\usepackage{amsmath}
\usepackage{amsthm}
\usepackage{amsfonts}
\usepackage{mathtools}
\usepackage{mathrsfs}
\usepackage{bm} 

\usepackage{graphicx}
\usepackage{dcolumn}
\usepackage[ruled,lined,algonl]{algorithm2e}
\usepackage{color}
\usepackage{autobreak}
\allowdisplaybreaks

\newcommand{\gai}[1]{{#1}}

\theoremstyle{plain}
\newtheorem{theorem}{Theorem}

\newtheorem{lemma}{Lemma}
\newtheorem{proposition}{Proposition}

\theoremstyle{definition}

\newtheorem{model}{Model}

\theoremstyle{remark}

\newcommand{\pset}[1]{\mathcal{#1}}

\DeclareMathOperator{\numer}{Num}
\DeclareMathOperator{\denom}{Den}
\DeclareMathOperator{\Det}{Det}

\DeclareMathOperator{\res}{res}

\usepackage{authblk}
\usepackage{listings}

\begin{document}
\begin{CJK}{UTF8}{gbsn}

\title{Influence of rationality levels on dynamics of heterogeneous Cournot duopolists with quadratic costs}


\author[a]{Xiaoliang Li}

\author[a]{Yihuo Jiang\thanks{Corresponding author: jiangyihuo@dgcu.edu.cn}}

\affil[a]{School of Digital Economics, Dongguan City University, Dongguan 523419, China}


\date{}
\maketitle

\begin{abstract}
This paper is intended to investigate the dynamics of heterogeneous Cournot duopoly games, where the first players adopt identical gradient adjustment mechanisms but the second players are endowed with distinct rationality levels. Based on tools of symbolic computations, we introduce a new approach and use it to establish rigorous conditions of the local stability for these models. We analytically investigate the bifurcations and prove that the period-doubling bifurcation is the only possible bifurcation that may occur for all the considered models. The most important finding of our study is regarding the influence of players' rational levels on the stability of heterogeneous duopolistic competition. It is derived that the stability region of the model where the second firm is rational is the smallest, while that of the one where the second firm is boundedly rational is the largest. This fact is counterintuitive and contrasts with relative conclusions in the existing literature. Furthermore, we also provide numerical simulations to demonstrate the emergence of complex dynamics such as periodic solutions with different orders and strange attractors.

Key Words: rationality level; heterogeneous duopoly; isoelastic demand; quadratic cost; symbolic computation
\end{abstract}

\section{Introduction}

In perfect competition, a large number of small companies compete with each other. In contrast, we use the term ``duopoly'' to denote a market where two firms coexist, which is, however, much more complicated than a competitive one. This is because players in duopolistic competition are influenced by their rivals' reactions when making decisions. It is well known that the first formal theory of oligopoly was developed by Cournot \cite{Cournot1838R}, where firms were supposed to make choices on their output quantity and have perfect information regarding their rivals' strategies. Cournot's model can be formulated by a linear map due to the linearity of the market demand and firm costs. The unique equilibrium is globally stable in the case of duopolistic competition. However, the model will be destabilized as the number of firms increases. For example, Theocharis \cite{Theocharis1960O} derived that in Cournot's game the equilibrium may become unstable when more than three players are involved.

Afterward, many authors contributed to the strand of introducing nonlinearity to either the demand or cost function. Fisher \cite{Fisher1961T} indicated diseconomies of scale (quadratic costs) can improve the stability of oligopolistic competition with boundedly rational players. Still, the demand function of Fisher's model is linear. However, through various empirical validations in the literature, we may reject the linear market demand function because nonlinearity was widely observed. Accordingly, Puu \cite{Puu1991C} introduced a discrete Cournot duopoly model with a nonlinear demand curve and demonstrated that complex dynamics can emerge easily. 

Puu's work inspired many follow-ups and modifications. For example, Kopel \cite{Kopel1996S} analyzed a coupled logistic duopoly model that can give rise to multiple equilibria, and showed the possibility of coexisting periodic and strange attractors. Ahmed et al.\ \cite{Ahmed2000O} extended Puu's model to that with boundedly rational firms producing differentiated goods, and indicated that the local stability can be improved by introducing delays. Bischi et al.\ \cite{Bischi2007O} considered a Cournot oligopoly, where players possess incomplete information regarding the market demand but conjecture prices using the Local Monopolistic Approximation (LMA) mechanism. They found the counterintuitive fact that less information may lead to more stability. Furthermore, Naimzada and Tramontana \cite{Naimzada2009C} proved the global stability of the model studied by Bischi et al. \cite{Bischi2007O} if only two firms are involved. Elsadany \cite{Elsadany2012C} considered a triopoly game where three boundedly rational players compete with each other, and discovered that chaotic behaviors may occur as the adjustment speeds increase. Cavalli et al.\ \cite{Cavalli2015N} studied a heterogeneous nonlinear duopoly with an LMA player and a gradient adjustment player, where multiple stable attractors coexist. Other related work includes \cite{Canovas2018O, Hommes2013B, Wu2010C, Ma2013T, Matsumoto2015D, Matouk2017N, Agiza1998E, Askar2016N, Elsadany2017D, Li2022C, Li2020N, Baiardi2018A, Agliari2016N}, etc.

It is interesting to explore the influence of rationality levels on the dynamics of heterogeneous Cournot competition. To address this issue, Cavalli and Naimzada \cite{Cavalli2015Na} compared several dynamic duopoly games, which are differentiated in terms of players' rationality degrees. They considered rational players with perfect foresight, boundedly rational players with naive expectations, and reduced rationality players with LMA adjustment mechanisms when the market possesses an isoelastic demand curve and the firms have linear cost functions. Motivated by \cite{Cavalli2015Na}, this study also focuses on the influence of rationality levels but extends to the case of quadratic costs or decreasing returns to scale. 

More specifically, we consider a market with the same demand function as \cite{Cavalli2015Na}, but extend the case of linear costs to that of quadratic costs. We investigate the dynamics of three Cournot duopoly models, named GR, GB, and GL, respectively. In these models, the first firms identically adopt the gradient adjustment mechanism but the second players are endowed with distinct rationality levels. We employ several tools of symbolic computations including, e.g., the triangular decomposition, resultant, and \gai{PCAD method}, in our analytical investigations. It is worth noting that the results of symbolic computations are exact, and thus can provide theoretical foundations for systematic analysis of economic models \cite{Huang2019A}. Based on symbolic computations, we introduce a new approach and use it to establish rigorous conditions of the local stability for Models GR and GL for the first time. We analyze the local bifurcations and derive that the period-doubling bifurcation is the only possible bifurcation that may emerge for all the models. The most important finding of this work is that the stability region of Model GR is the smallest, while that of Model GB is the largest. This fact is counterintuitive and contrasts with relative results in the existing literature, e.g., \cite{Naimzada2009C, Bischi2007O, Cavalli2015Na}. We conjecture that the involvement of the gradient adjustment mechanism plays an ambitious role in the emergence of these surprising phenomena. Furthermore, to demonstrate the complex dynamics of Models GR and GL, we conduct numerical simulations, through which periodic solutions with different orders and strange attractors can be observed.

The remainder of this article is organized as follows. In Section 2, we construct duopoly games endowed with different rationality levels. In Section 3, the local stability and bifurcations of the models are analytically investigated. Section 4 explores the influence of rational levels on the size of the stability region of heterogeneous duopolistic competition. In Section 5, we provide numerical simulations to demonstrate the existence of complex dynamics. Concluding remarks are provided in Section 6.

\section{Models}

Assume that the market is supplied by two firms producing homogeneous products. Our study focuses on heterogeneous duopolistic competition, where players adjust their output according to different dynamic mechanisms.

In what follows, $x_i(t)$ represents the output quantity of the $i$-th firm at period $t$. We assume that $C_i(x_i)=c_i x_i^2$ with $c_i>0$, meaning the cost function of the $i$-th firm is quadratic. Furthermore, the demand function of the market is supposed to be isoelastic. Specifically, the price function or the inverse demand function is
$$p(S)=\frac{1}{S}=\frac{1}{x_1+x_2},$$
where $S=x_1+x_2$ is the aggregate production of the commodity. 

Assume that period $t+1$ is the current period and period $t$ is the last period. In each model considered in this paper, we always suppose that the first firm adjusts its output quantity according to a \emph{gradient adjustment} mechanism, in which information regarding the latest marginal profit is needed. In detail, the profit of the first firm in the last period is
$$\Pi_1(t)=\frac{x_1(t)}{x_1(t)+x_{2}(t)}-c_1x_1^2(t),$$
where $x_{2}(t)$ is the output of its rival, namely the second firm. Then, at the current period, firm 1 decides its output according to 
\begin{equation}
	x_1(t+1) = x_1(t) + \delta x_1(t) \frac{\partial \Pi_1(t)}{\partial x_1(t)},
\end{equation}
where 
$$\frac{\partial \Pi_1(t)}{\partial x_1(t)}=\frac{x_{2}(t)}{(x_1(t)+x_{2}(t))^2}-2\,c_1x_1(t)$$ 
is the marginal profit and $\delta>0$ is a parameter. One can see that the adjustment speed is controlled by the parameter $\delta$. For the sake of simplicity, we denote 
$$G_1(x_1(t),x_{2}(t))=x_1(t)\frac{\partial \Pi_1(t)}{\partial x_1(t)}=\frac{x_1(t)x_{2}(t)}{(x_1(t)+x_{2}(t))^2}-2\,c_1x_1^2(t).$$

In each model, the second player is supposed to adjust its strategies based on the expectation of profits. Specifically, at the current period $t+1$, the second firm expects its profit to be
$$\Pi_2^e(t+1)=p_2^e(t+1) x_2(t+1) - c_2x_2^2(t+1),$$
where $p_2^e(t+1)$ is the estimation by firm 2 of the possible product price at the current period. To maximize the expected profit, firm 2 chooses to produce
$$x_2(t+1)=\arg\max_{x_2(t+1)}\Pi_2^e(t+1)=\arg\max_{x_2(t+1)}\left[p_2^e(t+1) x_2(t+1) - c_2x_2^2(t+1)\right].$$

The formulation of $p_2^e(t+1)$ is dependent on the rationality level of the firm. First, we consider the case that the second firm is a \emph{rational} player that has complete information regarding the inverse demand function and the production plan of its rival. On the one hand, if firm 2 knows the form of the inverse demand function, then it will expect the product price to be 
$$p_2^e(t+1)=\frac{1}{x_2(t+1)+x_1^e(t+1)},$$
where $x_{1}^e(t+1)$ is the estimation of firm 1's output. On the other hand, if firm 2 possesses the production plan of its rival, then it does not need to estimate the output of firm 1. That is to say, we can simply assume that $x_{1}^e(t+1)=x_{1}(t+1)$. Therefore, the expected profit of firm 2 is equal to the realized one, i.e.,
$$\Pi_2^e(t+1)=\Pi_2(t+1)=\frac{x_2(t+1)}{x_2(t+1)+x_{1}(t+1)}-c_2x_2^2(t+1).$$

To maximize the profit function above, we consider the first-order condition
\begin{equation}\label{eq:rational-cd-r}
x_{1}(t+1)-2\,c_2x_2(t+1)(x_2(t+1)+x_{1}(t+1))^2=0,
\end{equation}
which is a cubic polynomial equation and denoted as $F_2(x_2(t+1),x_{1}(t+1))= 0$. The second player can maximize its profit by solving $x_2(t+1)$ from this equation. Only one positive solution exists, namely
\begin{equation}\label{eq:r_i}
	x_2(t+1)=\frac{\sqrt[3]{2}M}{6\,c_2}+\frac{\sqrt[3]{4}\,c_2x_{1}^2(t+1)}{3\,M}-\frac{2\,x_{1}(t+1)}{3},
\end{equation}
where
$$M=\sqrt[3]{c_2^2	x_{1}(t+1)(4\,c_2x_{1}^2(t+1)+3\sqrt{3}\sqrt{8\,c_2x_{1}^2(t+1)+27}+27)}.$$

For simplicity, we denote \eqref{eq:r_i} by 
$$x_2(t+1)=R_2(x_{1}(t+1)).$$ 
In short, we name this game Model GR and formulate it as the following iteration map.

\begin{model}[GR]
\begin{equation}\label{eq:sys-gr}
	M_{GR}(x_1,x_2): 
	\left\{\begin{split}
		&x_1(t+1)=x_1(t)+\delta G_1(x_1(t),x_2(t)),\\
		&x_2(t+1)=R_2(x_1(t+1)),
	\end{split}
	\right.
\end{equation}
where $\delta>0$.
\end{model}

The assumption of a rational player is too strong in general. In the real world, however, a firm can hardly collect information on business secrets such as the production plan of its competitor. In the second model, we assume that firm 2 is boundedly rational rather than completely rational. Similar to a rational player, a \emph{boundedly rational} player is also capable to possess the exact form of the nonlinear demand function. However, the player's rationality is bounded in the sense that it has no idea about other's competing strategies but conjectures its rival to produce the same output as the last period. Specifically, if firm 2 is boundedly rational, then it naively expects firm 1's output to be $x_{1}^e(t+1)=x_{1}(t)$. Accordingly, the expected profit is 
$$\Pi_2^e(t+1)=\frac{x_2(t+1)}{x_2(t+1)+x_{1}(t)}-c_2x_2^2(t+1).$$
Therefore, $x_2(t+1)=R_2(x_{1}(t))$ according to the principle of maximum profit. We have the following model.

\begin{model}[GB]
	\begin{equation}\label{eq:sys-gb}
	M_{GB}(x_1,x_2): 
	\left\{\begin{split}
		&x_1(t+1)=x_1(t)+\delta G_1(x_1(t),x_2(t)),\\
		&x_2(t+1)=R_2(x_1(t)),
	\end{split}
	\right.
\end{equation}
where $\delta>0$.
\end{model}

Model GB has been intensively studied in \cite{Li2022A}. However, its formulation is still given here because we wish to compare it with the other two models with distinct rationality degrees. 

The last agent we consider has a further reduced rationality degree and is called a \emph{local monopolistic approximation} (LMA) player. Compared to a boundedly rational player, an LMA player even do not possess the formulation of the inverse demand function. However, it is realistic to assume an LMA player can observe the price of the commodity and the corresponding aggregate production from the last period. Moreover, operation skills such as market surveys and business experiments permit an LMA firm to estimate the slope of the demand function at the last period. Therefore, an LMA firm is able to conjecture the commodity price at the current period using the formula:
$$p_2^e(t+1)=p(S(t))+p'(S(t))(x_2^e(t+1)-S(t)),$$
where $x_2^e(t+1)=x_2(t+1)+x_{1}^e(t+1)$ represents the aggregate production expected by firm 2. 

For the sake of simplicity, we also suppose that an LMA player adopts the naive expectations, meaning $x_{1}^e(t+1)=x_{1}(t)$. In consequence,  
$$p_2^e(t+1)=\frac{1}{S(t)}-\frac{1}{S^2(t)}(x_2(t+1)-x_2(t)).$$
Accordingly, the expected profit is
$$\Pi^e_2(t+1)=x_2(t+1)\left[\frac{1}{S(t)}-\frac{1}{S^2(t)}(x_2(t+1)-x_2(t))\right]-c_2x_2^2(t+1).$$
To maximize this expected profit, one can solve the first-order condition and obtain
$$x_2(t+1)=\frac{2\,x_2(t)+x_{1}(t)}{2\left(1+c_2(x_2(t)+x_{1}(t))^2\right)}.$$
We simply denote the formula above as 
$$x_2(t+1)=S_2(x_2(t),x_{1}(t)).$$

Therefore, the third model can be formulated by the following iteration map.

\begin{model}[GL]
\begin{equation}\label{eq:sys-gl}
	M_{GL}(x_1,x_2): 
	\left\{\begin{split}
		&x_1(t+1)=x_1(t)+\delta G_1(x_1(t),x_2(t)),\\
		&x_2(t+1)=S_2(x_1(t),x_2(t)),
	\end{split}
	\right.
\end{equation}	
where $\delta>0$.
\end{model}

\section{Analysis of Local Stability and Bifurcations}

This section aims at analyzing the stability and bifurcations of the equilibrium for Models GR and GL. In \cite{Li2022A}, Li and Su have thoroughly explored the stability conditions for Model GB by analytically solving the equilibrium and plugging it into the Jury criterion \cite{Jury1976I}. However, in the sequel, we introduce a different approach based on symbolic computations to compute the rigorous conditions of the local stability without knowing in advance the closed-form equilibrium. 

\subsection{Model GR}

Map \eqref{eq:sys-gr} can be equivalently described by a one-dimensional iteration map, i.e.,
\begin{equation}\label{eq:gr-map-dim1}
	x_1(t+1)=x_1(t)+\delta G_1(x_1(t),R_2(x_1(t))).
\end{equation}
By setting $x_1(t+1)=x_1(t)=x_1$, we acquire the equilibrium equation
$$x_1=x_1+\delta G_1(x_1,R_2(x_1)),$$
which is equivalent to 
\begin{equation*}
	\begin{split}
	\left\{\begin{split}
		& G_1(x_1,x_2)=0,\\
		&F_2(x_2,x_1)=0,
	\end{split}
	\right.		
	\end{split}
\end{equation*}
i.e.,
\begin{equation*}
	\begin{split}
	\left\{\begin{split}
		& \frac{x_1x_{2}}{(x_1+x_{2})^2}-2\,c_1x_1^2=0,\\
		&x_{1}-2\,c_2x_2(x_2+x_{1})^2=0.
	\end{split}
	\right.		
	\end{split}
\end{equation*}

The solutions of this system can be decomposed into the two triangular sets in \eqref{eq:triset} using the method of triangular decomposition. This method can be used to transform a polynomial equation into triangular forms without changing any solutions. Readers may refer to \cite{Wu1986B, Kalkbrener1993A, Aubry1999T, Wang2000C, Li2010D} for more information.
\begin{equation}\label{eq:triset}
\begin{split}
	&\pset{T}_{1}=[x_2,x_1],\\
	&\pset{T}_{2}=\left[(4\,c_1c_2x_1^2 - c_2)x_2 + (2\,c_1^2 + 2\,c_1c_2)x_1^3, ~(4\,c_1^3 - 8\,c_1^2c_2 + 4\,c_1c_2^2)x_1^4 + 8\,c_1c_2x_1^2 - c_2\right].		
\end{split}
\end{equation}
The zero of $\pset{T}_1$ is the origin $(0,0)$, which should be ignored since $G_1$ or the iteration map \eqref{eq:sys-gr} is not defined on it. From $\pset{T}_2$, only one positive equilibrium can be obtained:
\begin{equation}\label{eq:equilibrium}
(x_1,x_2)=\left(~\frac{\sqrt{c_2}}{\sqrt{c_1}+\sqrt{c_2}}\frac{1}{\sqrt{2\sqrt{c_1c_2}}},~ \frac{\sqrt{c_1}}{\sqrt{c_1}+\sqrt{c_2}}\frac{1}{\sqrt{2\sqrt{c_1c_2}}}~\right).	
\end{equation}

Therefore, there is one unique equilibrium for Model GR, which is the Nash equilibrium. To identify whether this equilibrium is locally stable, we consider the derivative
\begin{equation}\label{eq:deriv-gr}
DQ\equiv \frac{{\rm d}x_1(t+1)}{{\rm d}x_1(t)}\bigg|_{x_1(t)=x_1,x_2(t)=x_2} =1+\delta \frac{\partial G_1(x_1,x_2)}{\partial x_1} +\delta \frac{\partial G_1(x_1,x_2)}{\partial x_2}\frac{\textrm{d} x_2}{\textrm{d} x_1},	
\end{equation}
where $\frac{\textrm{d} x_2}{\textrm{d} x_1}=\frac{\textrm{d} R_2}{\textrm{d} x_1}$ might not be obtained directly but can be calculated using the implicit function differentiation. According to \eqref{eq:rational-cd-r}, we have
$$x_{1}-2\,c_2 R_2(x_1) (R_2(x_1)+x_{1})^2=0.$$
Taking the derivative of this equation with respect to $x_1$ and solving $\frac{\textrm{d} R_2}{\textrm{d} x_1}$, we have
$$\frac{{\rm d} R_2}{{\rm d} x_1}=-\frac{4\,c_2x_1x_2+4\,c_2x_2^2-1}{2\,c_2(x_1^2+4\,x_1x_2+3\,x_2^2)}.$$

It is known that an equilibrium $(x_1,x_2)$ of map \eqref{eq:sys-gr} is locally stable provided that $|DQ|<1$. Furthermore, according to classical bifurcation theory, the system may undergo a period-doubling (flip) or fold bifurcation if the equilibrium loses its stability at $DQ=-1$ or $DQ=1$, respectively. One obvious way to establish the conditions of the two bifurcations is plugging the analytical expression of the Nash equilibrium \eqref{eq:equilibrium} into \eqref{eq:deriv-gr}. However, in the sequel, we introduce another approach based on the resultant, which is more general and does not require knowing in advance the closed-form equilibrium.

Suppose that we have two
univariate polynomials in $x$ with coefficients $a_i,b_j$, and $a_m,b_l\neq 0$, namely
$$A=\sum_{i=0}^ma_i\,x^i,~~B=\sum_{j=0}^lb_j\,x^j.$$
The determinant
\begin{equation*}\label{eq:sylmat}
 \begin{array}{c@{\hspace{-5pt}}l}
 \left|\begin{array}{cccccc}
a_m & a_{m-1}& \cdots   & a_0   &        &       \\
           & \ddots   & \ddots&    \ddots    &\ddots&   \\
         &          & a_m   & a_{m-1}&\cdots& a_0 \\ [5pt]
b_l & b_{l-1}& \cdots   &  b_0 &    &         \\
            & \ddots   &\ddots &   \ddots     &\ddots&       \\
         &   &    b_{l}     & b_{l-1} & \cdots &  b_0
\end{array}\right|
& \begin{array}{l}\left.\rule{0mm}{8mm}\right\}l\\
\\\left.\rule{0mm}{8mm}\right\}m
\end{array}
\end{array}
\end{equation*}
is called the \emph{Sylvester resultant} (or simply
\emph{resultant}) of $A$ and $B$ with respect to $x$, and denoted by $\res(A,B,x)$. 	The following lemma (refer to, e.g., \cite{Mishra1993A}) describes the main property of the resultant.

\begin{lemma}\label{lem:res-com}
Let $A$ and $B$ be two univariate polynomials in $x$.
Two polynomials $F$ and $G$ exist such that 
$$FA+GB=\res(A,B,x).$$
Furthermore, $\res(A, B)=0$ is equivalent to that $A$ and $B$ have common zeros in the field of complex numbers.
\end{lemma}

For a triangular set $\pset{T}=[T_1(x),T_2(x,y)]$ and a polynomial $H(x,y)$, we define 
$$\res(H,\pset{T})\equiv \res(\res(H,T_2,y),T_1(x),x).$$
By Lemma \ref{lem:res-com}, if $T_1=0$ and $T_2=0$ (or simply denoted as $\pset{T}=0$), one knows that $H=0$ implies $\res(H,\pset{T})=0$, which means $\res(H,\pset{T})=0$ is a necessary condition of $H=0$. It should be noted that the resultant is feasible only for polynomials. In this paper, $\numer(\cdot)$ and $\denom(\cdot)$ stand for the numerator and denominator, respectively, of a rational function. Concerning the condition of the fold bifurcation, i.e., $1-DQ=0$, we consider its numerator $\numer(1-DQ)$ and have that
$$\res(\numer(1-DQ),\pset{T}_{2})=68719476736\, c_2^{20} c_1^{14} \left(c_1-c_2\right)^{10} \left(c_1+c_2\right)^{20} \delta ^{4}.$$
Let $\pset{T}_{2}=0$.
As $c_1>0$, $c_2>0$, and $\delta >0$, we know that $\res(\numer(1-DQ),\pset{T}_{2})=0$ if $c_1-c_2=0$. In the following discussion, readers will see that the Nash equilibrium does not lose its stability at $c_1-c_2=0$, meaning that there are no fold bifurcations for the equilibrium of Model GR. Concerning the period-doubling bifurcation, we have
$$\res(\numer(1+DQ),\pset{T}_{2})=1073741824\, c_2^{19} c_1^{11} \left(c_1-c_2\right)^{10} \left(c_1+c_2\right)^{20} R_{GR},$$
where
$$R_{GR}=64\,c_1^3c_2\delta ^4-96\,c_1^2c_2\delta ^2-81\,c_1^2+18\,c_1c_2-c_2^2.$$
Then, $\res(\numer(1+DQ),\pset{T}_{2})$ vanishes if $c_1-c_2=0$ or $R_1=0$. As aforementioned, the Nash equilibrium does not lose the local stability at $c_1-c_2=0$. Consequently, the system may undergo a period-doubling bifurcation only when $R_1=0$.

To investigate the local stability, one needs to identify the conditions on the parameters that $1-DQ>0$ and $1+DQ>0$ are both fulfilled. Obviously, the signs of $1-DQ$ and $1+DQ$ are the same as those of $\numer(1-DQ)\cdot\denom(1-DQ)$ and $\numer(1+DQ)\cdot\denom(1+DQ)$, respectively, provided that $\denom(1-DQ)\neq 0$ and $\denom(1+DQ)\neq 0$. One can compute that
\begin{align*}
&\res(\numer(1-DQ)\cdot\denom(1-DQ),\pset{T}_{2})=\\
&\quad -1152921504606846976\, c_2^{39} c_1^{24} \left(c_1-c_2\right)^{18} \left(c_1+c_2\right)^{40} \left(9\, c_1-c_2\right)^{2} \delta ^{4},\\	
&\res(\numer(1+DQ)\cdot\denom(1+DQ),\pset{T}_{2})=\\
&\quad -18014398509481984\, c_2^{38} c_1^{21} \left(c_1-c_2\right)^{18} \left(c_1+c_2\right)^{40} \left(9\, c_1-c_2\right)^{2} R_{GR}.
\end{align*}

The signs of $\numer(1-DQ)\cdot\denom(1-DQ)$ and $\res(\numer(1-DQ)\cdot\denom(1-DQ),\pset{T}_{2})$ may be different. However, it should be emphasized that $\res(\numer(1-DQ)\cdot\denom(1-DQ),\pset{T}_{2})=0$ divides the parameter set $\{(c_1,c_2,\delta )\,|\,c_1> 0, c_2>0, \delta >0\}$ into separated regions and in each region the sign of $1-DQ$ is invariant. For $1+DQ$, we have a similar fact. Consequently, we can just pick one sample point from each of the regions divided by $\res(\numer(1-DQ)\cdot\denom(1-DQ),\pset{T}_{2})=0$ and $\res(\numer(1+DQ)\cdot\denom(1+DQ),\pset{T}_{2})=0$, and determine the signs of $1-DQ$ and $1+DQ$ at these sample points. 

The selection of sample points might be quite complex in general and can be automatically conducted using systematic algorithms based on symbolic computations. \gai{For example, e.g., the method of partial cylindrical algebraic decomposition (PCAD) \cite{Collins1991P}} permits us to select at least one sample point from each of the regions divided by several algebraic varieties (zeros of polynomials). It is worth noting that the computational results produced by this algorithm are exact and can be used to prove mathematical theorems rigorously. Readers can refer to \cite{Li2014C} for additional details. By using the \gai{PCAD method}, we find 6 sample points, which are listed in Table \ref{tab:sample-gr}. The information on the stability of the equilibrium, i.e., whether $1-DQ >0$ and $1+DQ>0$ are simultaneously satisfied at each sample point, is also reported. Furthermore, Table \ref{tab:sample-gr} shows the sign of $R_{GR}$ on every sample point. One can see that for any sample point, Model GR is stable if $R_{GR}<0$, and vice versa. That is to say, $R_{GR}<0$ is a necessary and sufficient condition of the local stability for Model GR. Therefore, the equilibrium loses its stability only at $R_{GR}=0$ rather than at $c_1-c_2=0$. To summarize, Theorem \ref{thm:gr} is acquired.

\begin{table}[htbp]
	\centering 
	\caption{Selected Sample Points in $\{(c_1,c_2,\delta )\,|\,c_1> 0, c_2>0, \delta >0\}$ for Model GR}
	\label{tab:sample-gr} 
	\begin{tabular}{|l|c|c||l|c|c|}  
		\hline  
$(c_1,c_2,\delta )$ & stable & $R_{GR}$ & $(c_1,c_2,\delta )$ & stable & $R_{GR}$ \\ \hline
$(1, 1/2, 1)$ & yes & $-$ & $(1, 2, 2)$ & no & $+$ \\ \hline
$(1, 1/2, 2)$ & no & $+$ & $(1, 10, 1)$ & yes & $-$ \\ \hline
$(1, 2, 1)$ & yes & $-$ & $(1, 10, 2)$ & no & $+$ \\ \hline
	\end{tabular}
\end{table}

\begin{theorem}\label{thm:gr}
In Model GR, we have a unique equilibrium with $x_1,x_2>0$. This equilibrium is locally stable if $R_{GR}<0$, where 
$$R_{GR}=
64\,c_1^3c_2\delta ^4-96\,c_1^2c_2\delta ^2-81\,c_1^2+18\,c_1c_2-c_2^2.
$$
Furthermore, the only possible bifurcation for the equilibrium of Model GR is the period-doubling bifurcation that may take place when $R_{GR}=0$.
\end{theorem}

\begin{figure}[htbp]
  \centering
    \subfigure[$c_2=c_1$.]{\includegraphics[width=0.4\textwidth]{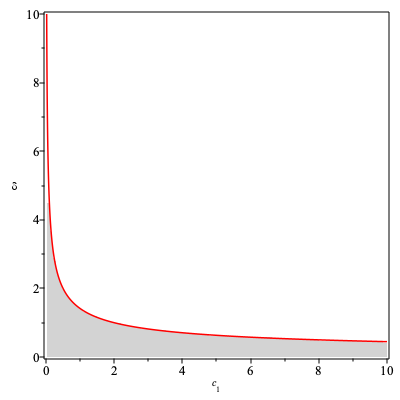}} 
    \subfigure[$\delta =1$.]{\includegraphics[width=0.4\textwidth]{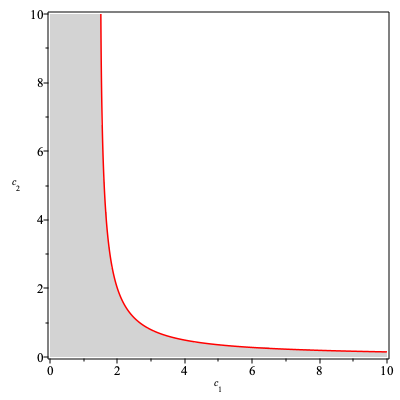}}
  \caption{Cross-sections of the stability region of Model GR with one parameter fixed. The stability regions are marked in grey.}
    \label{fig:par-gr}
\end{figure}

In Figure \ref{fig:par-gr}, we depict two cross-sections of the stability region of Model GR with one parameter fixed. The stability regions are marked in grey. It can be observed that the equilibrium might lose its local stability if any of $c_1$, $c_2$, $\delta $ is large enough.

\subsection{Model GL}

The equilibrium equations of map \eqref{eq:sys-gl} are
\begin{equation*}
	\left\{\begin{split}
		& G_1(x_1,x_2)=0,\\
		&S_2(x_2,x_1)=0,\\
	\end{split}
	\right.
\end{equation*}
of which the solutions can be decomposed into zeros of the same two triangular sets as in \eqref{eq:triset}. Accordingly, we derive that there exists one unique positive equilibrium and its closed-form expression is \eqref{eq:equilibrium}. 

The Jacobian matrix of map \eqref{eq:sys-gl} can be written as
\begin{equation*}
J=\left[
	\begin{matrix}
	J_{11} & J_{12}\\
	J_{21} & J_{22}	
	\end{matrix}
\right]=
\left[
	\begin{matrix}
	1+\delta \cdot \partial G_1/\partial x_1 & \delta \cdot \partial G_1/\partial x_2\\
	\partial S_2/\partial x_1 & \partial S_2/\partial x_2
	\end{matrix}
\right].
\end{equation*}
Then, the characteristic polynomial of $J$ is
$$CP(\lambda)=\lambda^2-(J_{11}+J_{22})\lambda+\Det(J),$$
where $\Det(J)=J_{11}J_{22}-J_{12}J_{21}$ is the determinant of $J$.
The local stability analysis can be conducted by investigating whether all the eigenvalues of $J$, i.e., all the roots of $CP$, are in the open unit disk. To the best of our knowledge, several criteria are feasible for this purpose. Here, we employ the Jury criterion \cite{Jury1976I}, which indicates that the equilibrium is locally stable if the following conditions are fulfilled:
\begin{enumerate}
	\item $CD_1^J\equiv CP(1)>0$,
	\item $CD_2^J\equiv CP(-1)>0$,
	\item $CD_3^J\equiv 1-\Det(J)>0$.
\end{enumerate}
According to bifurcation theory, for a discrete dynamic system, when $CD_1^J=0$, $CD_2^J=0$, or $CD_3^J=0$, the equilibrium may undergo a fold, period-doubling, or Neimark-Sacker bifurcation, respectively.

We analyze the bifurcations by determining the possibility of $CD_i^J=0$. At the equilibrium, it is known that $\res(\numer(CD_i^J),\pset{T}_2)=0$ is a necessary condition of $CD_i^J=0$. Concerning the fold bifurcation, we have
$$\res(\numer(CD_1^J),\pset{T}_2)=17592186044416\,c_2^{28} c_1^{18} \left(4\, c_1 -c_2\right)^{2} \left(c_1-c_2\right)^{14} \left(c_1+c_2\right)^{28} \delta ^{4}.$$
Let $\pset{T}_2=0$. As $c_1,c_2,\delta >0$, we know that $\res(\numer(CD_1^J),\pset{T}_2)=0$ if $c_1-c_2=0$ or $4\,c_1-c_2=0$. To explore the period-doubling bifurcation, we compute
$$\res(\numer(CD_2^J),\pset{T}_2)=274877906944\, c_2^{25} c_1^{15} \left(4\, c_1 -c_2\right)^{2} \left(c_1-c_2\right)^{12} \left(c_1+c_2\right)^{28} R_{GL}^1,$$
where
\begin{align*}
	R_{GL}^1=\,&
64\,c_1^7c_2\delta ^4-672\,c_1^6c_2^2\delta ^4+1796\,c_1^5c_2^3\delta ^4-168\,c_1^4c_2^4\delta ^4+4\,c_1^3c_2^5\delta ^4+384\,c_1^6c_2\delta ^2\\
&-400\,c_1^5c_2^2\delta ^2-2136\,c_1^4c_2^3\delta ^2+96\,c_1^3c_2^4\delta ^2+8\,c_1^2c_2^5\delta ^2-256\,c_1^6+544\,c_1^5c_2\\
&-353\,c_1^4c_2^2+100\,c_1^3c_2^3-38\,c_1^2c_2^4+4\,c_1c_2^5-c_2^6.
\end{align*}
Hence, $\res(\numer(CD_2^J),\pset{T}_2)=0$ happens if $c_1-c_2=0$, $4\,c_1-c_2=0$, or $R_{GL}^1=0$. Furthermore, we know
$$\res(\numer(CD_3^J),\pset{T}_2)=4294967296\, c_2^{26} c_1^{14} \left(4\, c_1 -c_2\right)^{2} \left(c_1-c_2\right)^{10} \left(c_1+c_2\right)^{28} R_{GL}^2,$$
where
\begin{align*}
	R_{GL}^2=\,&
64\, c_1^{7} \delta ^{4}-416\, c_1^{6} c_2 \delta ^{4}+708\, c_1^{5} c_2^{2} \delta ^{4}-104\, c_1^{4} c_2^{3} \delta ^{4}+4\, c_1^{3} c_2^{4} \delta ^{4}+48\, c_1^{5} c_2 \delta ^{2}+232\, c_1^{4} c_2^{2} \delta ^{2}\\
&-736\, c_1^{3} c_2^{3} \delta ^{2}-56\, c_1^{2} c_2^{4} \delta ^{2}-81\, c_1^{4} c_2+180\, c_1^{3} c_2^{2}-118\, c_1^{2} c_2^{3}+20\, c_1 c_2^{4}-c_2^{5}.
\end{align*}
Accordingly, $\res(\numer(CD_3^J),\pset{T}_2)$ vanish if $c_1-c_2=0$, $4\,c_1-c_2=0$, or $R_{GL}^2=0$. In the following discussion, one can see that the equilibrium loses its stability only at $R_{GL}^1=0$, which means that the only possible bifurcation for Model GL is the period-doubling bifurcation, which may occur when $R_{GL}^1=0$.

The stability conditions for Model GL are $CD_i^J>0$ ($i=1,2,3$), which are equivalent to $\numer(CD_i^J)\cdot \denom(CD_i^J)>0$ provided that $ \denom(CD_i^J)\neq 0$. The resultant of $\numer(CD_i^J)\cdot \denom(CD_i^J)$ with respect to $\pset{T}_2$ is as follows.
\begin{align*}
&\res(\numer(CD_1^J)\cdot\denom(CD_1^J),\pset{T}_2)=\\
&\quad -75557863725914323419136\, c_2^{53} c_1^{32} \left(c_1-c_2\right)^{28} \left(c_1+c_2\right)^{56} \left(4\, c_1-c_2\right)^{6} \delta ^{4},\\

&\res(\numer(CD_2^J)\cdot\denom(CD_2^J),\pset{T}_2)=\\
&\quad -1180591620717411303424\, c_2^{50} c_1^{29} \left(c_1-c_2\right)^{26} \left(c_1+c_2\right)^{56} \left(4\, c_1-c_2\right)^{6} R_{GL}^1,\\

&\res(\numer(CD_3^J)\cdot\denom(CD_3^J),\pset{T}_2)=\\
&\quad -18446744073709551616\, c_2^{51} c_1^{28} \left(c_1-c_2\right)^{24} \left(c_1+c_2\right)^{56} \left(4\, c_1-c_2\right)^{6} R_{GL}^2.
\end{align*}

It is noted that for each $i$ the sign of $\numer(CD_i^J)\cdot \denom(CD_i^J)$ may not be the same as that of $\res(\numer(CD_i^J)\cdot\denom(CD_i^J),\pset{T}_2)$ in general. However, the zeros of $\res(\numer(CD_i^J)\cdot\denom(CD_i^J),\pset{T}_2)$ divide the parameter set $\{(c_1,c_2,\delta )\,|\,c_1> 0, c_2>0, \delta >0\}$ into several regions, and in each region the sign of $\numer(CD_i^J)\cdot \denom(CD_i^J)$ or equivalently $CD_i^J$ is invariant. Hence, we can pick a sample point from every region, and determine whether the conditions $CD_i^J>0$ are fulfilled by checking them at these sample points. 

In Table \ref{tab:sample-gl}, we list 27 sample points selected by the \gai{PCAD method} and provide the information on whether the equilibrium is stable at these points, i.e., whether the stability conditions $CD_1^J>0$, $CD_2^J>0$, and $CD_3^J>0$ are simultaneously fulfilled. Furthermore, Table \ref{tab:sample-gl} reports the signs of $R_{GL}^1$ and $R_{GL}^2$ at these sample points. From Table \ref{tab:sample-gl}, one can see that the equilibrium is stable if and only if $R_{GL}^1<0$. In comparison, however, $R_{GL}^2<0$ is neither a necessary nor a sufficient condition of the local stability for Model GL. For example, at $(1, 1/8, 1)$, the equilibrium is stable but $R_{GL}^2>0$, which means that the stability of the equilibrium does not imply $R_{GL}^2<0$. Moreover, at $(1, 2, 3/2)$, we have $R_{GL}^2<0$ but the equilibrium is unstable, i.e., $R_{GL}^2<0$ does not imply the local stability.

\begin{table}[htbp]
	\centering 
	\caption{Selected Sample Points in $\{(c_1,c_2,\delta )\,|\,c_1> 0, c_2>0, \delta >0\}$ for Model GL}
	\label{tab:sample-gl} 
	\begin{tabular}{|l|c|c|c||l|c|c|c|}  
		\hline  
		$(c_1,c_2,\delta )$ & stable & $R_{GL}^1$ & $R_{GL}^2$ &$(c_1,c_2,\delta )$ &  stable & $R_{GL}^1$ & $R_{GL}^2$ \\ \hline
$(1, 1/8, 1/2)$& yes & $-$ & $-$ & $(1, 2, 3)$& no & $+$ & $+$ \\ \hline
$(1, 1/8, 1)$& yes & $-$ & $+$ & $(1, 5, 1)$& yes & $-$ & $-$ \\ \hline
$(1, 1/8, 3)$& no & $+$ & $+$ & $(1, 5, 2)$& no & $+$ & $-$ \\ \hline

$(1, 1/4, 1/2)$& yes & $-$ & $-$ & $(1, 5, 5)$& no & $+$ & $+$ \\ \hline
$(1, 1/4, 1)$& yes & $-$ & $+$ & $(1, 10, 1)$& yes & $-$ & $-$ \\ \hline
$(1, 1/4, 2)$& no & $+$ & $+$ & $(1, 10, 2)$& no & $+$ & $-$ \\ \hline

$(1, 9/32, 1/2)$& yes & $-$ & $-$ & $(1, 10, 22)$& no & $+$ & $+$ \\ \hline
$(1, 9/32, 1)$& yes & $-$ & $+$ & $(1, 13, 1)$& yes & $-$ & $-$ \\ \hline
$(1, 9/32, 2)$& no & $+$ & $+$ & $(1, 13, 2)$& no & $+$ & $-$ \\ \hline

$(1, 1/2, 1)$& yes & $-$ & $-$ & $(1, 13, 223)$& no & $+$ & $+$ \\ \hline
$(1, 1/2, 3/2)$& yes & $-$ & $+$ & $(1, 21, 1)$& yes & $-$ & $-$ \\ \hline
$(1, 1/2, 2)$& no & $+$ & $+$ & $(1, 21, 2)$& no & $+$ & $-$ \\ \hline

$(1, 2, 1)$& yes & $-$ & $-$ & $(1, 21, 13)$& no & $+$ & $+$ \\ \hline
$(1, 2, 3/2)$& no & $+$ & $-$ & &  &  &  \\ \hline
	\end{tabular}
\end{table}

\begin{theorem}\label{thm:gl}
In Model GL, we have a unique equilibrium with $x_1,x_2>0$. This equilibrium is locally stable if $R_{GL}^1<0$,
where 
\begin{align*}
	R_{GL}^1=\,&
64\,c_1^7c_2\delta ^4-672\,c_1^6c_2^2\delta ^4+1796\,c_1^5c_2^3\delta ^4-168\,c_1^4c_2^4\delta ^4+4\,c_1^3c_2^5\delta ^4+384\,c_1^6c_2\delta ^2\\
&-400\,c_1^5c_2^2\delta ^2-2136\,c_1^4c_2^3\delta ^2+96\,c_1^3c_2^4\delta ^2+8\,c_1^2c_2^5\delta ^2-256\,c_1^6+544\,c_1^5c_2\\
&-353\,c_1^4c_2^2+100\,c_1^3c_2^3-38\,c_1^2c_2^4+4\,c_1c_2^5-c_2^6.
\end{align*}
Furthermore, the only possible bifurcation for the equilibrium of Model GL is the period-doubling bifurcation that may occur when $R_{GL}^1=0$.
\end{theorem}

%

%

\begin{figure}[htbp]
  \centering
    \subfigure[$c_2=c_1$.]{\includegraphics[width=0.4\textwidth]{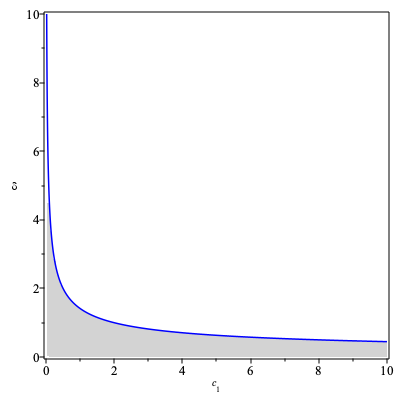}} 
    \subfigure[$\delta =1$.]{\includegraphics[width=0.4\textwidth]{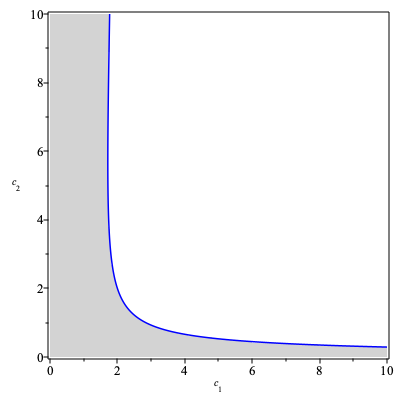}} \\  
    
    \caption{Cross-sections of the stability region of Model GL with one parameter fixed. The stability regions are marked in grey.}
    \label{fig:par-gl}
\end{figure}

Figure \ref{fig:par-gl} depicts two cross-sections of the stability region of Model GL reported in Theorem \ref{thm:gl}. One can see that the equilibrium remains stable if the values of $c_1$, $c_2$, and $\delta $ are sufficiently small.

\section{Comparison on Stability Regions}

This section concentrates on the influence of players' rationality levels on the stability of heterogeneous duopolist competition in face of diseconomies of scale. More specifically, we are concerned about the sizes of the stability regions of Models GR, GL, GB, and the inclusion relations among them. The condition of the local stability for Model GB was first reported in \cite[Theorem 2]{Li2022A}. Proposition \ref{prop:stable-gb} restates this theorem for the purpose of comparison. It should be mentioned that the stability condition of Proposition \ref{prop:stable-gb}, obtained by using the same approach as we derived Theorems \ref{thm:gr} and \ref{thm:gl}, is different but equivalent to that of \cite[Theorem 2]{Li2022A}.
 
\begin{proposition}\label{prop:stable-gb}
In Model GB, there is a unique equilibrium. This equilibrium is locally stable if $R_{GB}<0$,
where 
\begin{align*}
R_{GB}=&\,
4\,c_1^7c_2\delta ^4-272\,c_1^6c_2^2\delta ^4+4632\,c_1^5c_2^3\delta ^4-272\,c_1^4c_2^4\delta ^4+4\,c_1^3c_2^5\delta ^4+264\,c_1^6c_2\delta ^2\\
&-2464\,c_1^5c_2^2\delta ^2-6096\,c_1^4c_2^3\delta ^2+96\,c_1^3c_2^4\delta ^2+8\,c_1^2c_2^5\delta ^2-81\,c_1^6+342\,c_1^5c_2\\
&-559\,c_1^4c_2^2+436\,c_1^3c_2^3-159\,c_1^2c_2^4+22\,c_1c_2^5-c_2^6.
\end{align*}
Furthermore, the only possible bifurcation for the equilibrium of Model GB is the period-doubling bifurcation that may occur when $R_{GB}=0$.
\end{proposition}

\begin{figure}[htbp]
  \centering
    \subfigure[The 3-dimensional parameter space.]{\includegraphics[width=0.4\textwidth]{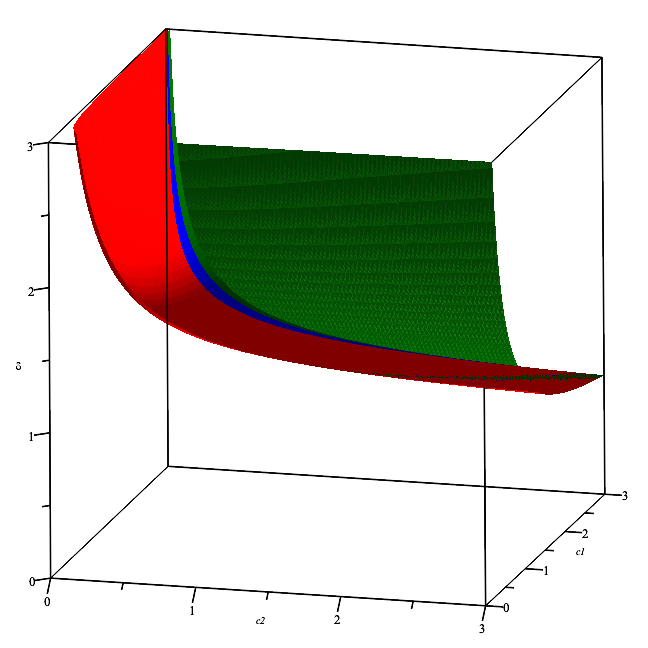}} 
    \subfigure[The cross-section with $\delta =1$.]{\includegraphics[width=0.4\textwidth]{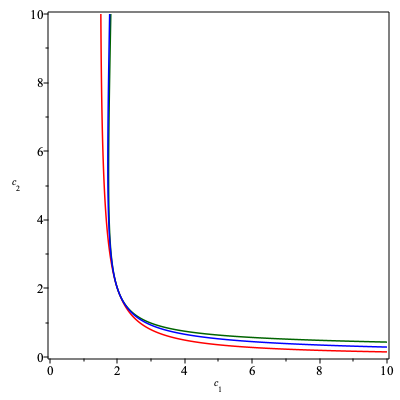}}
  \caption{The bifurcation surfaces (curves) of Models GR, GL, and GB, which are marked in red color, blue color, and green color, respectively.}
    \label{fig:compare}
\end{figure}

The only difference in the constructions of the three models is that the second firms are endowed with distinct rationality levels, i.e., a rational, boundedly rational, and LMA player in GR, GB, and GL, respectively. Figure \ref{fig:compare} presents the bifurcation surfaces (curves) $R_{GR}=0$, $R_{GL}^1=0$, and $R_{GB}=0$ of the three models. It seems that the region of the equilibrium stability for Model GR is the smallest, while that for Model GB is the largest. In the sequel, we turn these observations into facts by rigorously proving them with the aid of symbolic computations.

To investigate the inclusion relation between the stability regions of GR and GB, we employ again the \gai{PCAD method \cite{Collins1991P}}. The main idea is that $R_{GR}=0$ and $R_{GB}=0$ divide the parameter set $\{(c_1,c_2,\delta )\,|\,c_1> 0, c_2>0, \delta >0\}$ into serval regions. Evidently, in each region, the signs of $R_{GR}$ and $R_{GB}$ are invariant, which can be identified by checking them at a sample point selected by the \gai{PCAD method}. Then, we can conclude that $R_{GR}<0$ implies $R_{GB}<0$ or the stability region of GB covers that of GR if $R_{GB}<0$ is true on all the sample points where $R_{GR}<0$ is fulfilled.

In Table \ref{tab:grgl}, we list all the 16 sample points produced by the \gai{PCAD method} and provide the information on whether $R_{GR}<0$ and $R_{GB}<0$ are satisfied. It should be noted that the \gai{PCAD method} may generate multiple sample points for a single region rather than just one. In this case, however, the aforementioned approach still works. From Table \ref{tab:grgl}, one can see that $R_{GB}<0$ at all the sample points where $R_{GR}<0$. Therefore, we acquire Proposition \ref{prop:gr-gb}.

\begin{table}[htbp]
	\centering 
	\caption{Sample Points of $\{(c_1,c_2,\delta )\,|\,c_1> 0, c_2>0, \delta >0\}$ divided by $R_{GR}=0$ and $R_{GB}=0$}
	\label{tab:grgl} 
	\begin{tabular}{|l|c|c||l|c|c|}\hline  	
$(c_1,c_2,\delta )$  & $R_{GR}<0$ & $R_{GB}<0$ & $(c_1,c_2,\delta )$  & $R_{GR}<0$ & $R_{GB}<0$ \\ \hline
$(1/2, 1/2, 125/256)$  & true & true & $(1/2, 1/2, 141/128)$  & true & true \\ \hline
$(1/2, 1/2, 189/128)$  & true & true & $(1/2, 193/256, 115/64)$  & true & true \\ \hline
$(1/2, 477/64, 115/64)$  & false & true & $(1/2, 1779/128, 115/64)$  & true & true \\ \hline
$(1/2, 85/256, 495/256)$  & true & true & $(1/2, 173/256, 495/256)$  & false & true \\ \hline
$(1/2, 539/256, 495/256)$  & false & false & $(1/2, 4347/256, 495/256)$  & false & true \\ \hline
$(1/2, 495/16, 495/256)$  & true & true & $(1/2, 15/256, 5/2)$  & true & true \\ \hline
$(1/2, 275/2048, 5/2)$  & false & true & $(1/2, 4967/512, 5/2)$  & false & false \\ \hline
$(1/2, 24401/256, 5/2)$  & false & true & $(1/2, 22001/128, 5/2)$  & true & true\\ \hline
	\end{tabular}
\end{table}

\begin{proposition}\label{prop:gr-gb}
	The region of the local stability for Model GR is a proper subset of that for Model GB.
\end{proposition}

From Proposition \ref{prop:gr-gb}, we know that in face of quadratic costs, the stability region of heterogeneous Cournot duopolistic competition with a gradient adjustment player and a boundedly rational player will shrink if we replace the boundedly rational player with a completely rational player. This result is counterintuitive and surprises us. On the one hand, a homogenous oligopoly with rational players is always stable for all parameter values because the equilibrium would be achieved in one step if the involved firms know the strategies of their rivals clearly. In comparison, according to \cite{Puu1991C, Fisher1961T, Theocharis1960O}, homogenous oligopolies with boundedly rational players may be unstable as the number of players increases. In other words, the involvement of rational rather than boundedly rational players has an effect of stabilization on the dynamics of homogenous oligopolistic competition. On the other hand, under the assumptions of a nonlinear demand but linear costs, Cavalli et al.\ \cite{Cavalli2015Na} compared duopolistic competition with an LMA player and a boundedly rational player to that with an LMA player and a completely rational player. They derived that the existence of a completely rational player can improve stability. The finding of Proposition \ref{prop:gr-gb} contrasts with the two facts mentioned above in some sense.

The exploration of the reason for the above contradiction is important. As a comparison, we should mention that, in face of constant returns to scale, the stability region of duopolistic competition with a gradient adjustment player and a boundedly rational player will also shrink if we replace the boundedly rational firm with a rational one. Specifically, if $C_i(x_i)=c_ix_i$, one can derive that the heterogeneous duopolistic game with a gradient adjustment and a completely rational firm is locally stable if
\begin{equation}\label{eq:cx-gr}
	\delta (c_1+c_2) - 4<0.
\end{equation}
Furthermore, the game with the rational player replaced by a boundedly rational player has been thoroughly investigated by Tramontana \cite{Tramontana2010H}, where it is proved that the stability conditions include
\begin{equation}\label{eq:cx-gb}
\left\{
	\begin{split}
&\delta <\frac{4(c_1+c_2)}{4\,c_1c_2-(c_2-c_1)^2}\quad\text{if}~~0<c_1<\frac{1}{3}c_2~\text{or}~c_1\geq 3\,c_2,\\
&\delta <\frac{2(c_1+c_2)}{(c_2-c_1)^2}\quad\text{if}~~\frac{1}{3}c_2<c_1<3\,c_2.
	\end{split}
\right.
\end{equation}

One can derive that \eqref{eq:cx-gr} implies \eqref{eq:cx-gb}. Consequently, if the cost functions are linear, then the stability region will also shrink if replacing the boundedly rational company with a rational one. That is to say, the form of cost functions may not be the reason for the counterintuitive fact in Proposition \ref{prop:gr-gb}. We conjecture that the involvement of the gradient adjustment mechanism plays a crucial role herein.

Now, we compare the stability regions of Models GL and GB. By using the same approach as in Proposition \ref{prop:gr-gb}, we consider the parameter space divided by $R_{GL}=0$ and $R_{GB}=0$. The selected sample points and the corresponding information on whether the stability conditions $R_{GL}<0$ and $R_{GB}<0$ are satisfied are reported in Table \ref{tab:glgb}. One can see that $R_{GB}$ is negative on all the sample points where $R_{GL}$ is negative. Accordingly, Proposition \ref{prop:gl-gb} is obtained.

\begin{table}[htbp]
	
	\caption{Sample Points of $\{(c_1,c_2,\delta )\,|\,c_1> 0, c_2>0, \delta >0\}$ divided by $R_{GL}=0$ and $R_{GB}=0$}
	\label{tab:glgb} 
	\begin{tabular}{|l|c|c||l|c|c|}\hline  		
$(c_1,c_2,\delta )$  & $R_{GL}<0$ & $R_{GB}<0$ & $(c_1,c_2,\delta )$  & $R_{GL}<0$ & $R_{GB}<0$ \\ \hline
$(1/2, 15/2048, 689/256)$  & true & true & $(1/2, 15/2048, 759/128)$  & false & true \\ \hline
$(1/2, 15/2048, 893/128)$  & false & false & $(1/2, 393/16384, 499/256)$  & true & true \\ \hline
$(1/2, 393/16384, 1053/256)$  & false & true & $(1/2, 393/16384, 309/64)$  & false & false \\ \hline
$(1/2, 265/4096, 3/2)$  & true & true & $(1/2, 265/4096, 197/64)$  & false & true \\ \hline
$(1/2, 265/4096, 117/32)$  & false & false & $(1/2, 611/2048, 1105/1024)$  & true & true \\ \hline
$(1/2, 611/2048, 2213/1024)$  & false & true & $(1/2, 611/2048, 341/128])$  & false & false \\ \hline
$(1/2, 5/2, 239/256)$  & true & true & $(1/2, 5/2, 481/256)$  & false & true \\ \hline
$(1/2, 5/2, 153/64)$  & false & false & $(1/2, 1907/256, 133/128)$  & true & true \\ \hline
$(1/2, 1907/256, 537/256)$  & false & true & $(1/2, 1907/256, 335/128)$  & false & false \\ \hline
$(1/2, 1585/128, 145/128)$  & true & true & $(1/2, 1585/128, 73/32)$  & false & true \\ \hline
$(1/2, 1585/128, 179/64)$  & false & false & $(1/2, 4013/256, 303/256)$  & true & true \\ \hline
$(1/2, 4013/256, 611/256)$  & false & true & $(1/2, 4013/256, 93/32)$  & false &  false\\ \hline
$(1/2, 2239/128, 155/128)$  & true & true & $(1/2, 2239/128, 39/16)$  & false & true \\ \hline
$(1/2, 2239/128, 189/64)$  & false & false &   &  &  \\ \hline
\end{tabular}
\end{table}

\begin{proposition}\label{prop:gl-gb}
	The region of the local stability for Model GL is a proper subset of that for Model GB.
\end{proposition}

Proposition \ref{prop:gl-gb} is also contrary to the widely accepted fact \cite{Bischi2007O, Naimzada2009C, Cavalli2015Na} that the LMA mechanism leads to more stable trajectories compared to the boundedly rational one. In particular, Bischi et al.\ \cite{Bischi2007O} indicated that in an oligopoly game with homogeneous firms, for the LMA mechanism the equilibrium becomes unstable if at least five firms coexist in one market, while for the boundedly rational mechanism complex dynamics arise if there are four players. 

Similarly, we consider Models GR and GL. The selected sample points are listed in Table \ref{tab:grgl}, which implies Proposition \ref{prop:gr-gl}. To summarize, among the three models considered in our study, the smallest is the stability region of Model GR, while the largest is that of Model GB. These facts are counterintuitive and contrast with the current literature such as \cite{Naimzada2009C, Bischi2007O, Cavalli2015Na}. We reasonably conjecture that the involvement of the player adopting the gradient adjustment mechanism leads to these surprising results. However, it is still not clear why the dynamics are so different when a gradient adjustment player is involved in Cournot oligopolistic competition. We leave this problem for future research.

\begin{table}[htbp]
	\centering 
	\caption{Sample Points of $\{(c_1,c_2,\delta )\,|\,c_1> 0, c_2>0, \delta >0\}$ divided by $R_{GR}=0$ and $R_{GL}=0$}
	\label{tab:grgl} 
	
	\begin{tabular}{|l|c|c||l|c|c|}\hline  
$(c_1,c_2,\delta )$  & $R_{GR}<0$ & $R_{GL}<0$ & $(c_1,c_2,\delta )$  & $R_{GR}<0$ & $R_{GL}<0$ \\ \hline
$(1/2, 1/2, 117/256)$  & true & true & $(1/2, 1/2, 137/128)$  & true & true \\ \hline
$(1/2, 1/2, 189/128)$  & true & true & $(1/2, 199/256, 459/256)$  & true & true \\ \hline
$(1/2, 929/128, 459/256)$  & false & true & $(1/2, 1723/128, 459/256)$  & true & true \\ \hline
$(1/2, 43/128, 247/128)$  & true & true & $(1/2, 175/256, 247/128)$  & false & true \\ \hline
$(1/2, 595/256, 247/128)$  & false & false & $(1/2, 4331/256, 247/128)$  & false & true \\ \hline
$(1/2, 3889/128, 247/128)$  & true & true & $(1/2, 15/256, 5/2)$  & true & true \\ \hline
$(1/2, 265/2048, 5/2)$  & false & true & $(1/2, 10485/1024, 5/2)$  & false & false \\ \hline
$(1/2, 6135/64, 5/2)$  & false & true & $(1/2, 22001/128, 5/2)$  & true & true \\ \hline
	\end{tabular}
\end{table}

\begin{proposition}\label{prop:gr-gl}
	The region of the local stability for Model GR is a proper subset of that for Model GL.
\end{proposition}

\section{Numerical Simulations}

\begin{figure}[htbp]
  \centering
    \subfigure[$\delta \in (1.6,2.5)$.]{\includegraphics[width=0.45\textwidth]{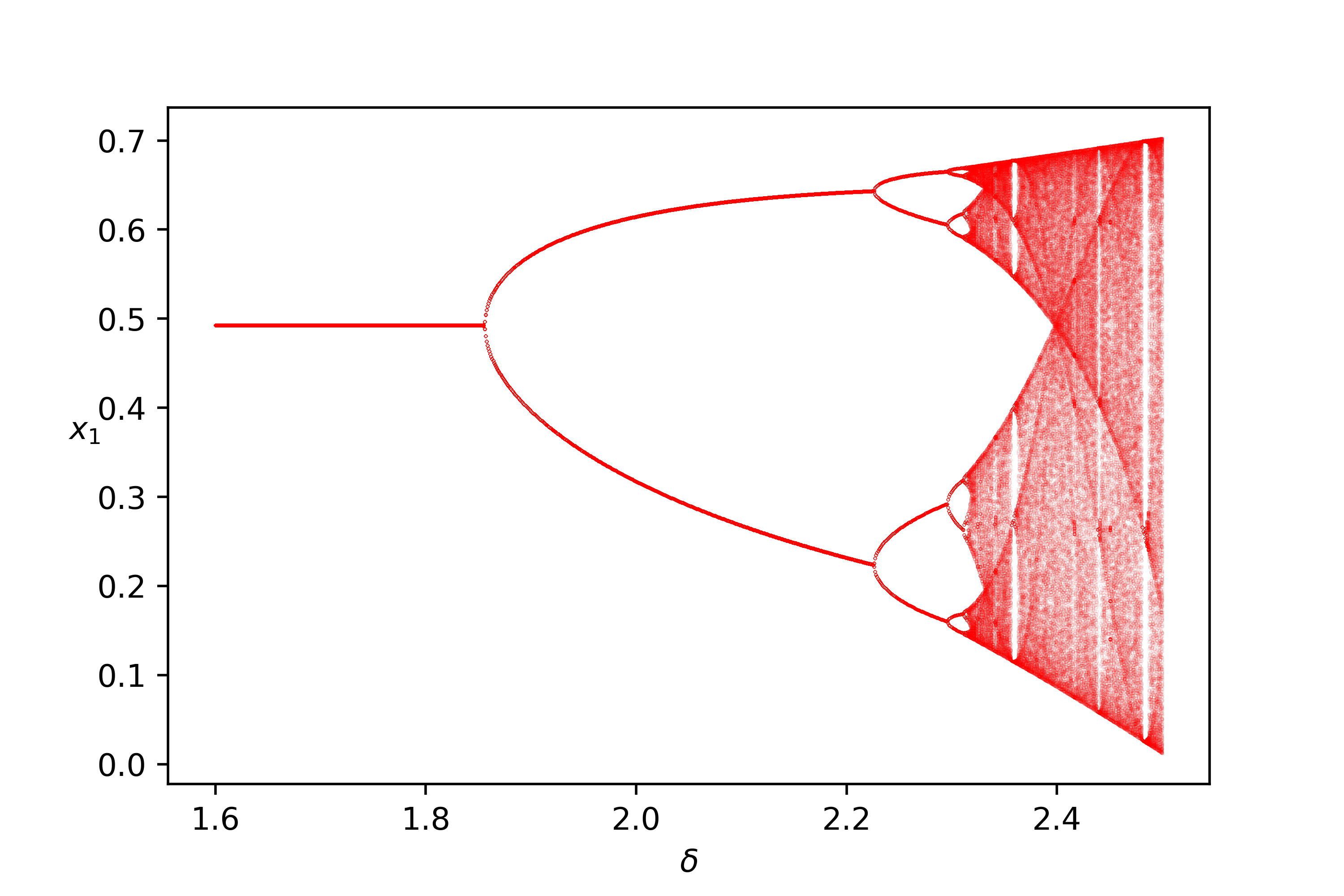}} 
    \subfigure[$\delta \in (2.45,2.5)$.]{\includegraphics[width=0.45\textwidth]{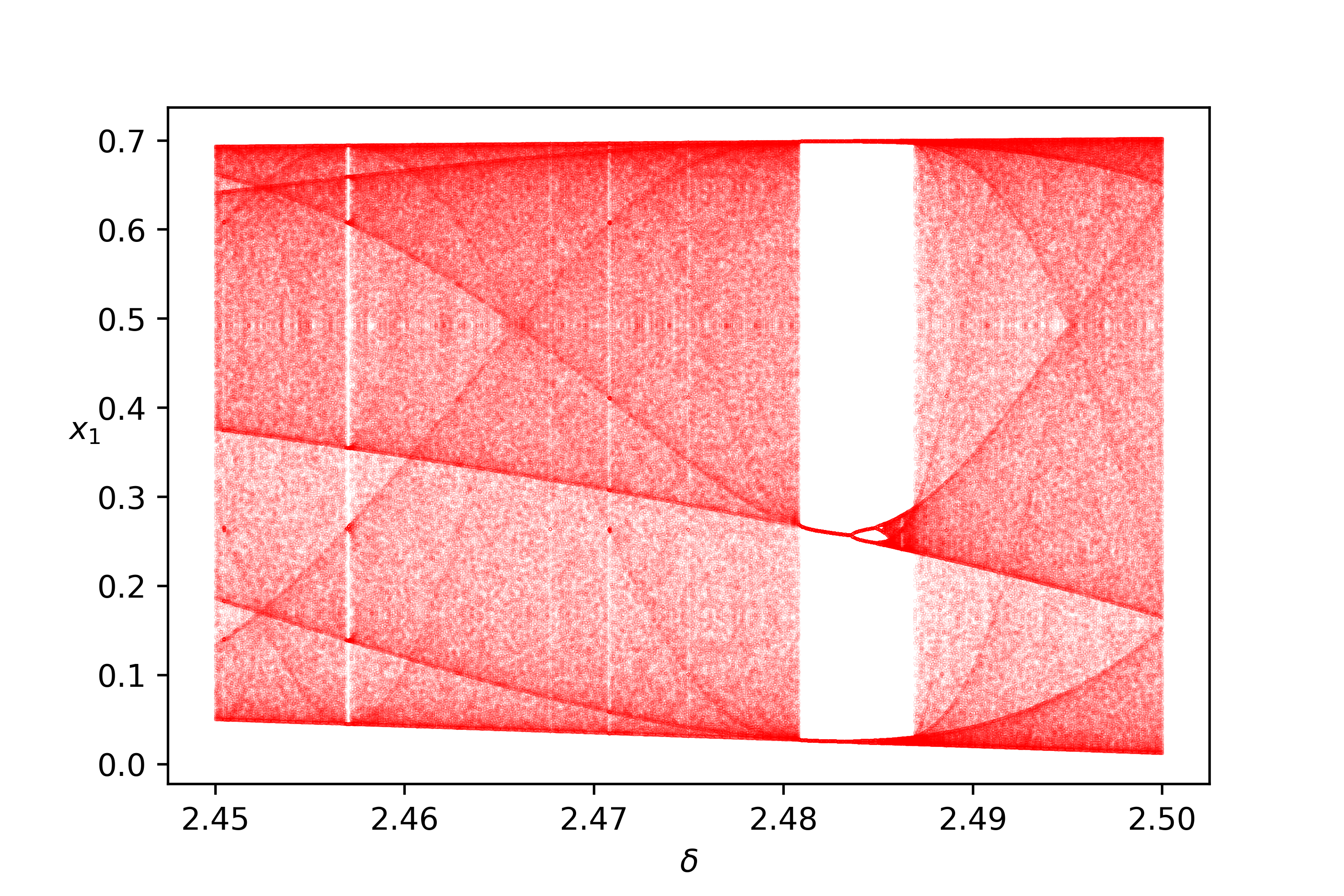}} \\
    
  \caption{The bifurcation diagrams of Model GR, i.e., map \eqref{eq:gr-map-dim1}, with respect to $\delta$ by fixing the parameters $c_1=0.5$, $c_2=1.0$. The diagram for $\delta \in(1.6,2.5)$ is provided in (a), and that for $\delta \in (2.45,2.5)$ is reported in (b). We choose $(x_1(0),x_2(0))=(0.5,0.5)$ to be the initial state of the iterations.}
    \label{fig:bif-gr}
\end{figure}

\begin{figure}[htbp]
  \centering
    \subfigure[Against $x_1$ for $\delta \in (1.3,2.5)$.]{\includegraphics[width=0.45\textwidth]{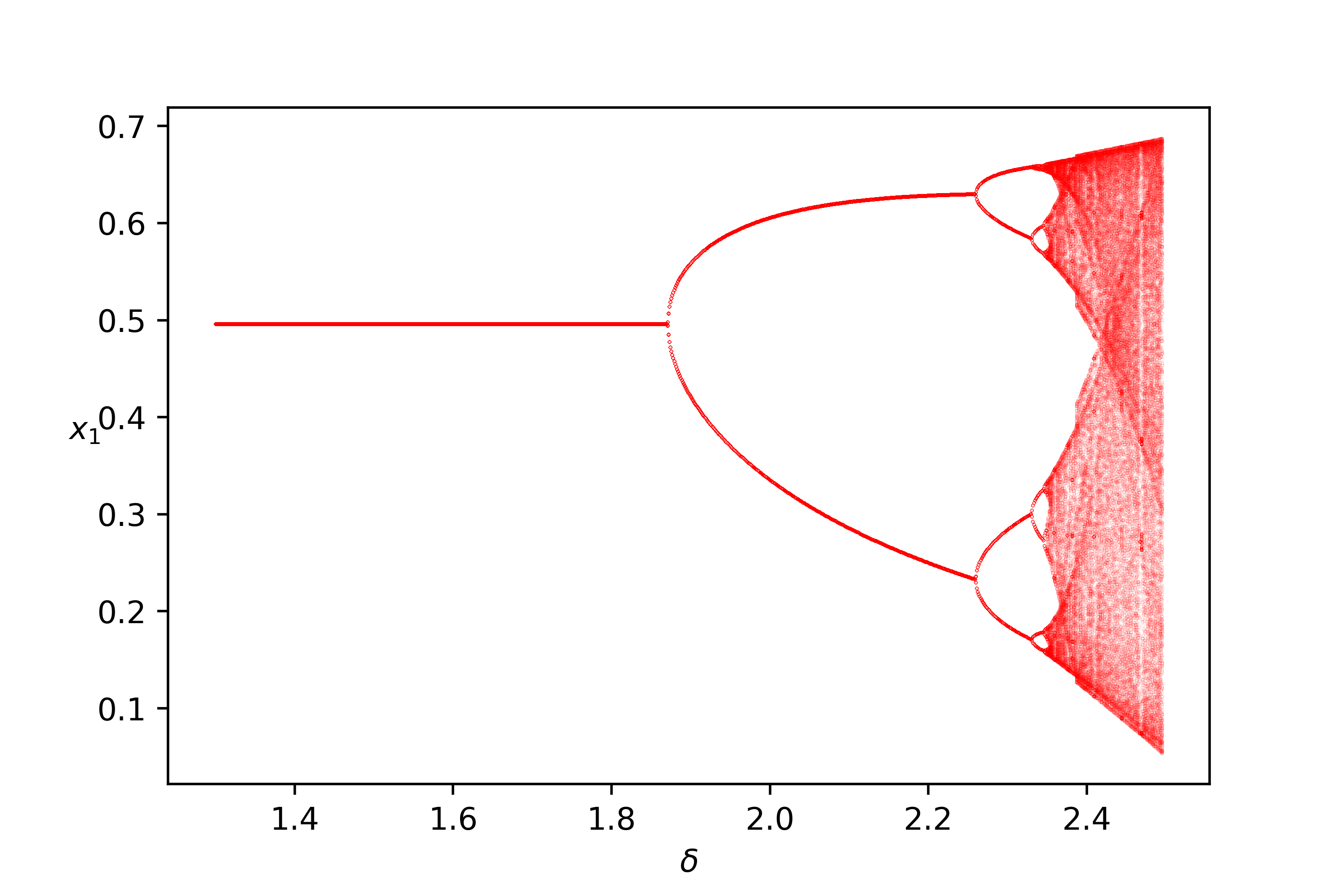}} 
    \subfigure[Against $x_2$ for $\delta \in (1.3,2.5)$.]{\includegraphics[width=0.45\textwidth]{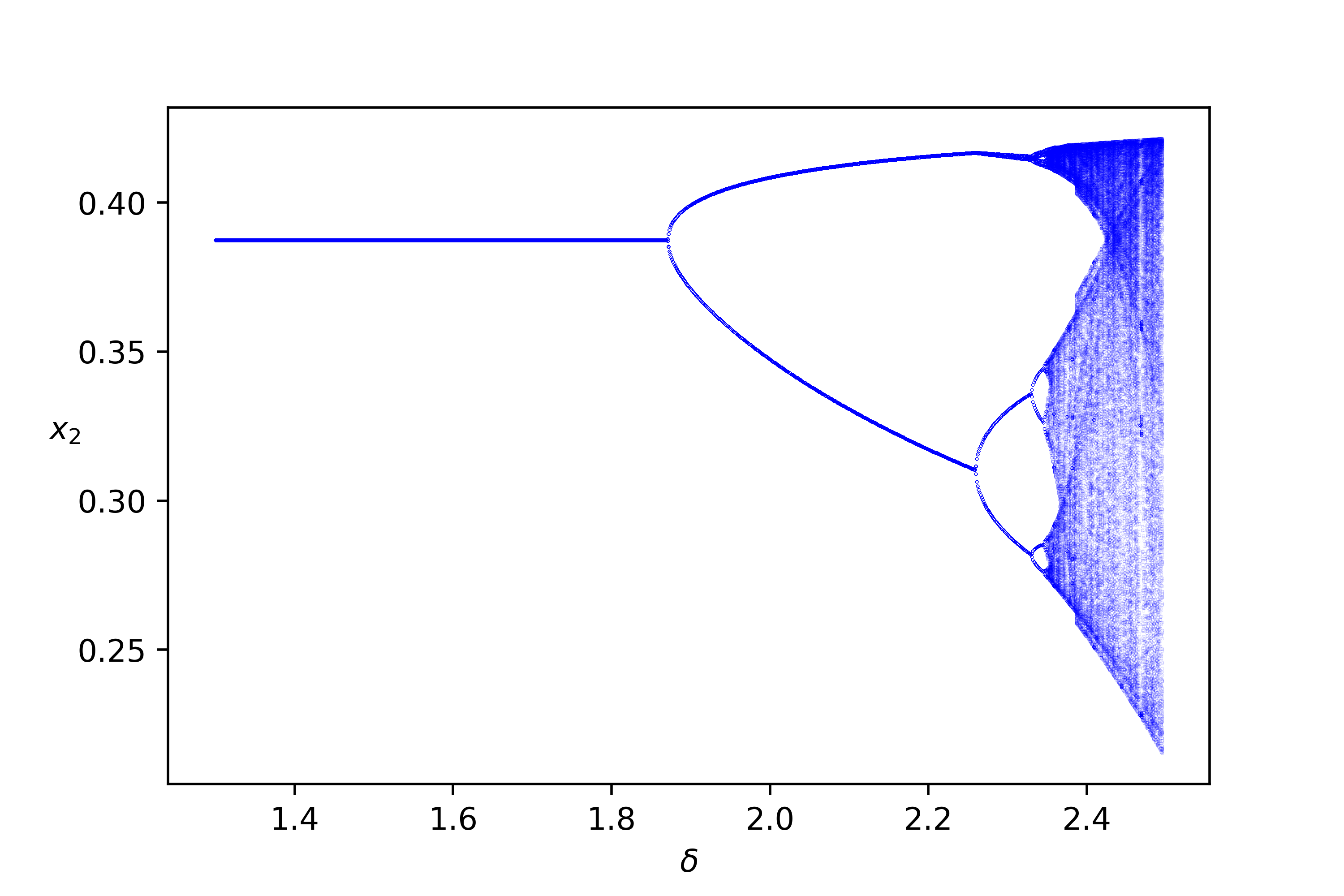}} \\
    \subfigure[Against $x_1$ for $\delta \in (2.45,2.5)$.]{\includegraphics[width=0.45\textwidth]{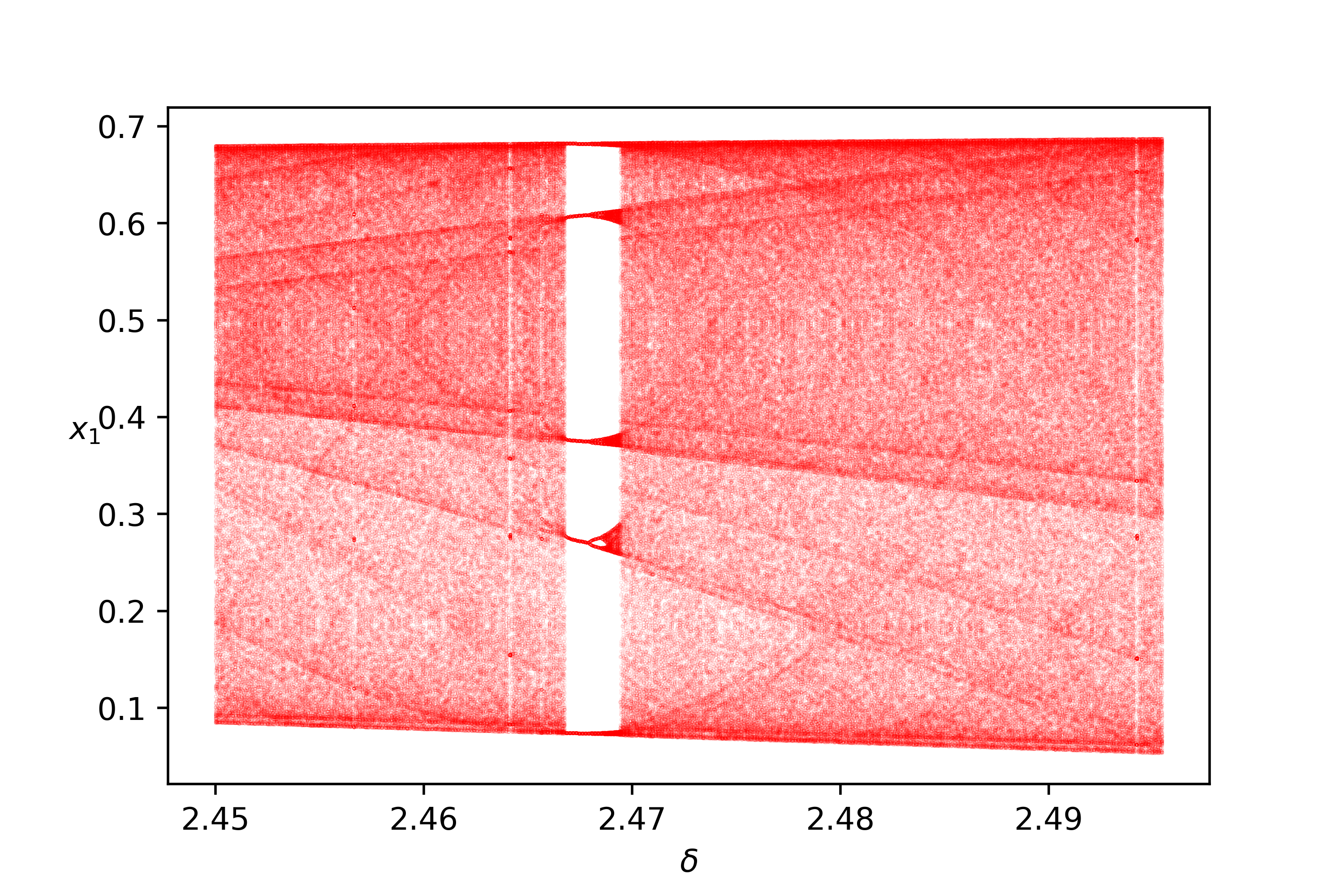}} 
    \subfigure[Against $x_2$ for $\delta \in (2.45,2.5)$.]{\includegraphics[width=0.45\textwidth]{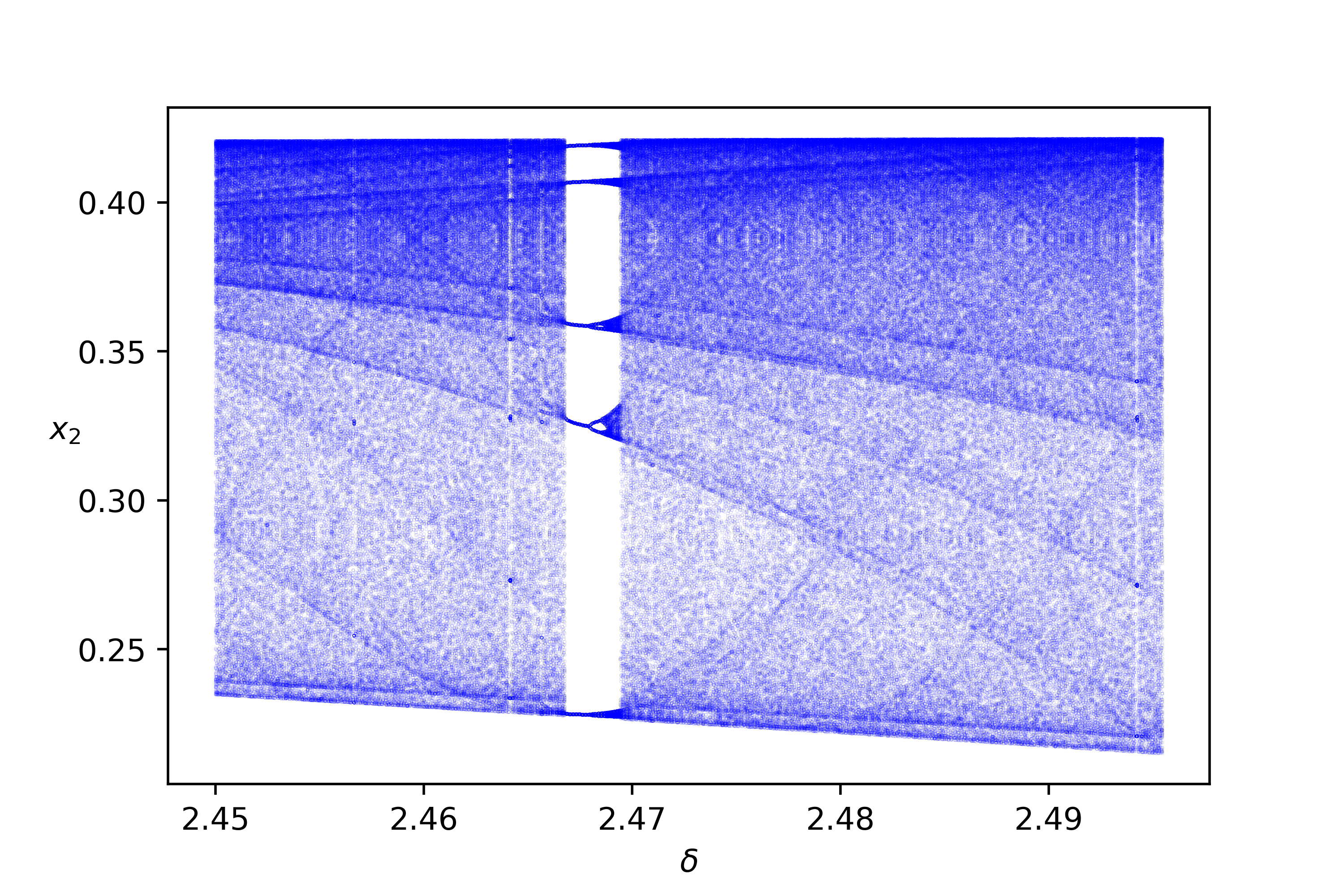}} \\
    
  \caption{The bifurcation diagrams of Model GL with respect to $\delta $ by fixing the parameters $c_1=0.5$, $c_2=1.0$. We choose $(x_1(0),x_2(0))=(0.5,0.5)$ to be the initial state of the iterations. For $\delta \in (1.3,2.5)$, the diagram against $x_1$ is given in (a), and that against $x_2$ is displayed in (b). Furthermore, for $\delta \in (2.45,2.5)$, the bifurcation diagram against $x_1$ is depicted in (c), and that against $x_2$ is provided in (d).}
    \label{fig:bif-gl}
\end{figure}

To illustrate the complex dynamics of Models GR and GL, numerical simulations are reported in this section, where one can observe periodic solutions with different orders and strange attractors. Figure \ref{fig:bif-gr} reports the bifurcation diagrams of the one-dimensional iteration map \eqref{eq:gr-map-dim1} (Model GR) with $c_1=0.5$ and $c_2=1.0$. The iterations start from the initial state $(x_1(0),x_2(0))=(0.5,0.5)$. The bifurcation diagrams for $\delta \in(1.6,2.5)$ and $\delta \in (2.45,2.5)$ are displayed in (a) and (b), respectively. One can see that the unique stable equilibrium loses its stability through a cascade of period-doubling bifurcations. In detail, the stable equilibrium bifurcates to a stable 2-cycle orbit at $\delta =1.855$, and then the 2-cycle orbit further bifurcates to a stable 4-cycle orbit at $\delta =2.227$. An 8-cycle orbit appears when $\delta =2.296$ and the dynamics of the model finally become chaotic when $\delta$ is sufficiently large. It is also found that period solutions with odd orders occur intermittently between the emergence of chaotic behaviors of the system. For example, Figure \ref{fig:bif-gl} (b) displays a stable 3-cycle orbit existing in a gap between the intervals of $\delta$ where chaos takes place. Furthermore, as the value of $\delta$ increases, this 3-cycle transitions into chaos again through a series of period-doubling bifurcations. It is well known that period three implies chaos in the sense of Li-Yorke for one-dimensional discrete dynamical systems \cite{Li1975P}. Consequently, our observations confirm that in Model GR chaos can arise for an uncountable set of initial states.

\begin{figure}[htbp]
  \centering
	\includegraphics[width=0.9\textwidth]{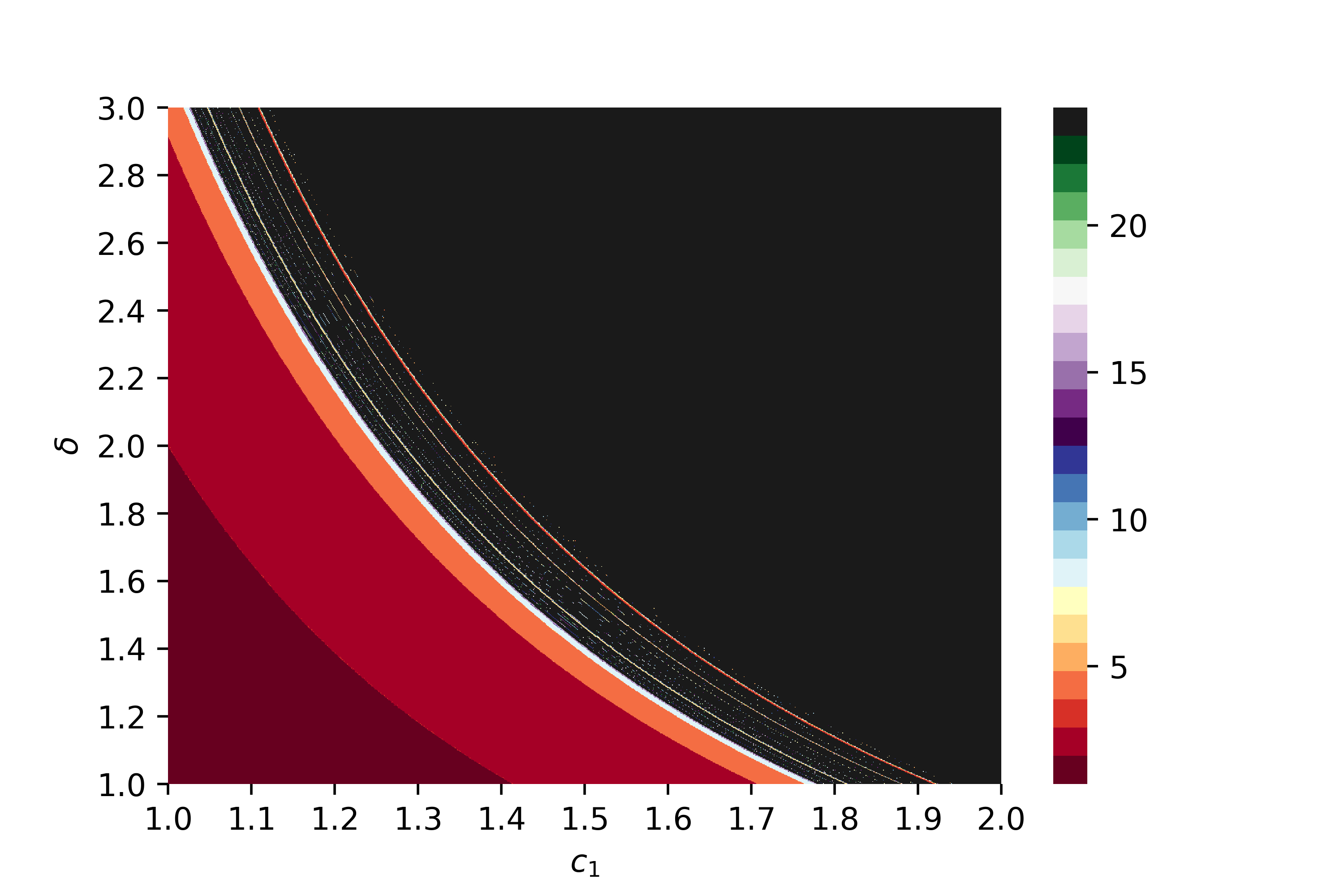}
  \caption{The 2-dimensional bifurcation diagram of Model GR, i.e., map \eqref{eq:gr-map-dim1}, for $(c_1,\delta)\in[1.0,2.0]\times[1.0,3.0]$ by fixing the parameter $c_2=c_1$. We choose $(x_1(0),x_2(0))=(0.1,0.1)$ to be the initial state of the iterations.}
    \label{fig:2d-bif-gr}
\end{figure}

\begin{figure}[htbp]
  \centering
	\includegraphics[width=0.9\textwidth]{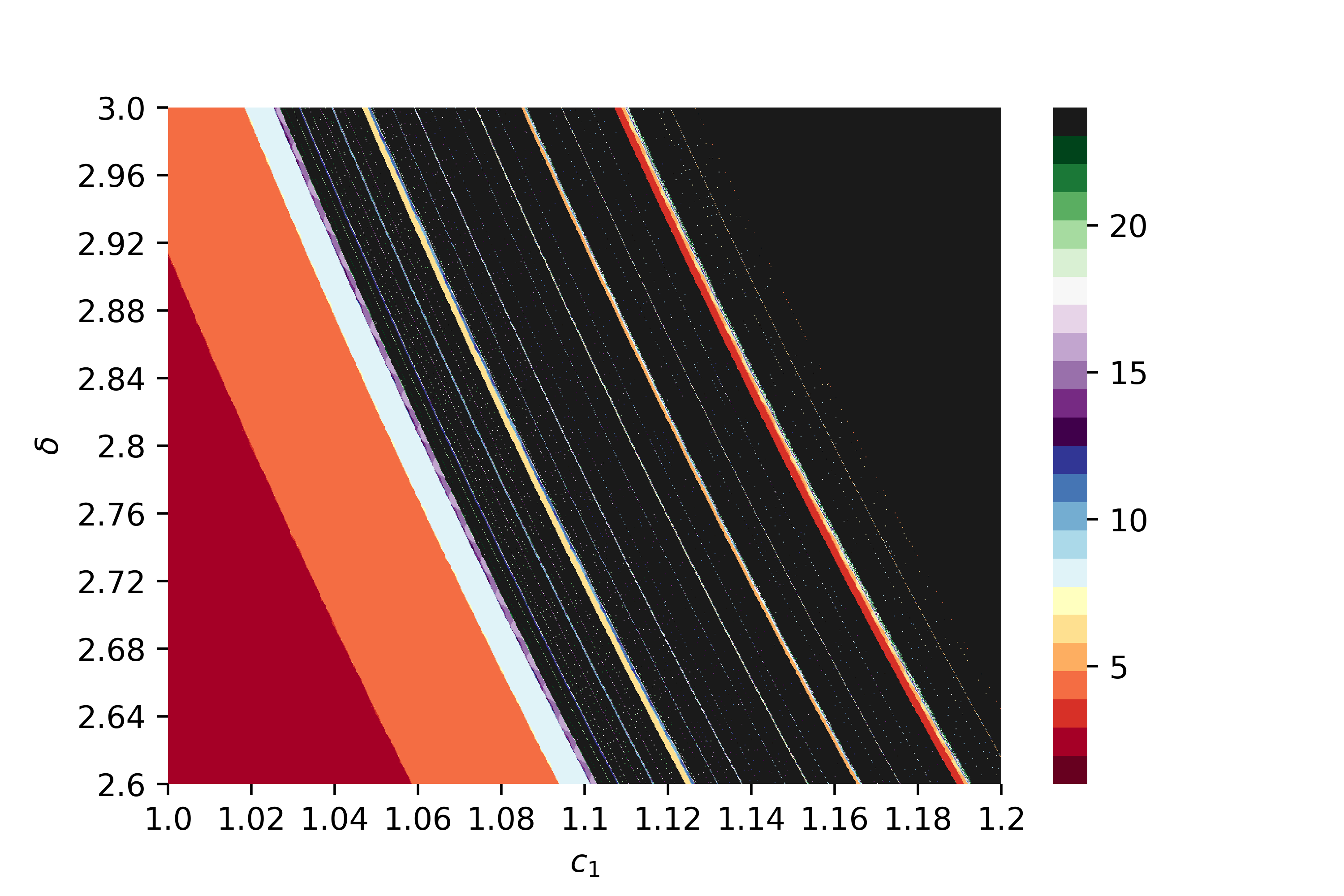}
  \caption{The 2-dimensional bifurcation diagram of Model GR, i.e., map \eqref{eq:gr-map-dim1}, for $(c_1,\delta)\in[1.0,1.2]\times[2.6,3.0]$ by fixing the parameter $c_2=c_1$. We choose $(x_1(0),x_2(0))=(0.1,0.1)$ to be the initial state of the iterations.}
    \label{fig:2d-bif-gr-fine}
\end{figure}

More complex dynamic behaviors of Model GR can be observed in the two-dimensional bifurcation diagrams presented in Figures \ref{fig:2d-bif-gr} and \ref{fig:2d-bif-gr-fine}. Readers can refer to \cite{Li2022C} for more details regarding two-dimensional bifurcation diagrams. In numerical simulations of producing these bifurcation diagrams, we fix the parameter $c_2=c_1$ and set the initial state as $(x_1(0),x_2(0))=(0.1,0.1)$. Parameter points corresponding to periodic orbits with different orders are marked in different colors and are marked in black if the order is greater than $24$. The black points may be viewed as the parameter values where complex dynamics such as chaos take place. The transitions between different types of periodic orbits can be observed in Figure \ref{fig:2d-bif-gr}. One can see that the equilibrium loses its stability through a series of period-doubling bifurcations as the value of $c_1$ or $\delta$ increases. Furthermore, periodic orbits seem to appear in gaps between chaotic dynamics. Additional details regarding the transitions can be found in Figure \ref{fig:2d-bif-gr-fine}, where an enlarged diagram is provided. One can see more clearly that a narrow parameter region corresponding to 3-cycles (marked in orange) is surrounded by the regions corresponding to chaos (marked in black). This implies that in Model GR chaos exists in the sense of Li-Yorke according to \cite{Li1975P}.

In Figure \ref{fig:bif-gl}, we present the bifurcation diagrams of map \eqref{eq:sys-gl} (Model GL), where the parameters are fixed as $c_1=0.5$ and $c_2=1.0$. The trajectories against $x_1$ and $x_2$ are colored in red and blue, respectively. Similar to Model GR, we find that the dynamics of Model GL transition from one unique stable equilibrium to strange attractors also through a series of period-doubling bifurcations. The unique stable equilibrium bifurcates into a stable 2-cycle orbit at $\delta =1.870$, which further bifurcates into a 4-cycle orbit at $\delta =2.259$. There is a stable 8-cycle orbit when $\delta =2.330$. Finally, chaotic dynamics take place if the value of $\delta$ is large enough. Enlarged bifurcation diagrams for $\delta \in (2.45,2.5)$ are given in (c, d), where a 5-cycle orbit can be discovered. As $\delta$ increases, this 5-cycle orbit transitions into chaos once more also through period-doubling bifurcations. It should be emphasized that the equilibria of Models GR and GL lose their local stability at $\delta =1.855$ and $\delta =1.870$, respectively. This confirms the conclusion in Proposition \ref{prop:gr-gl} that the region of the local stability for Model GR is covered by that for Model GL.

\begin{figure}[htbp]
  \centering
    \subfigure[$\delta =2.300000$.]{\includegraphics[width=0.4\textwidth]{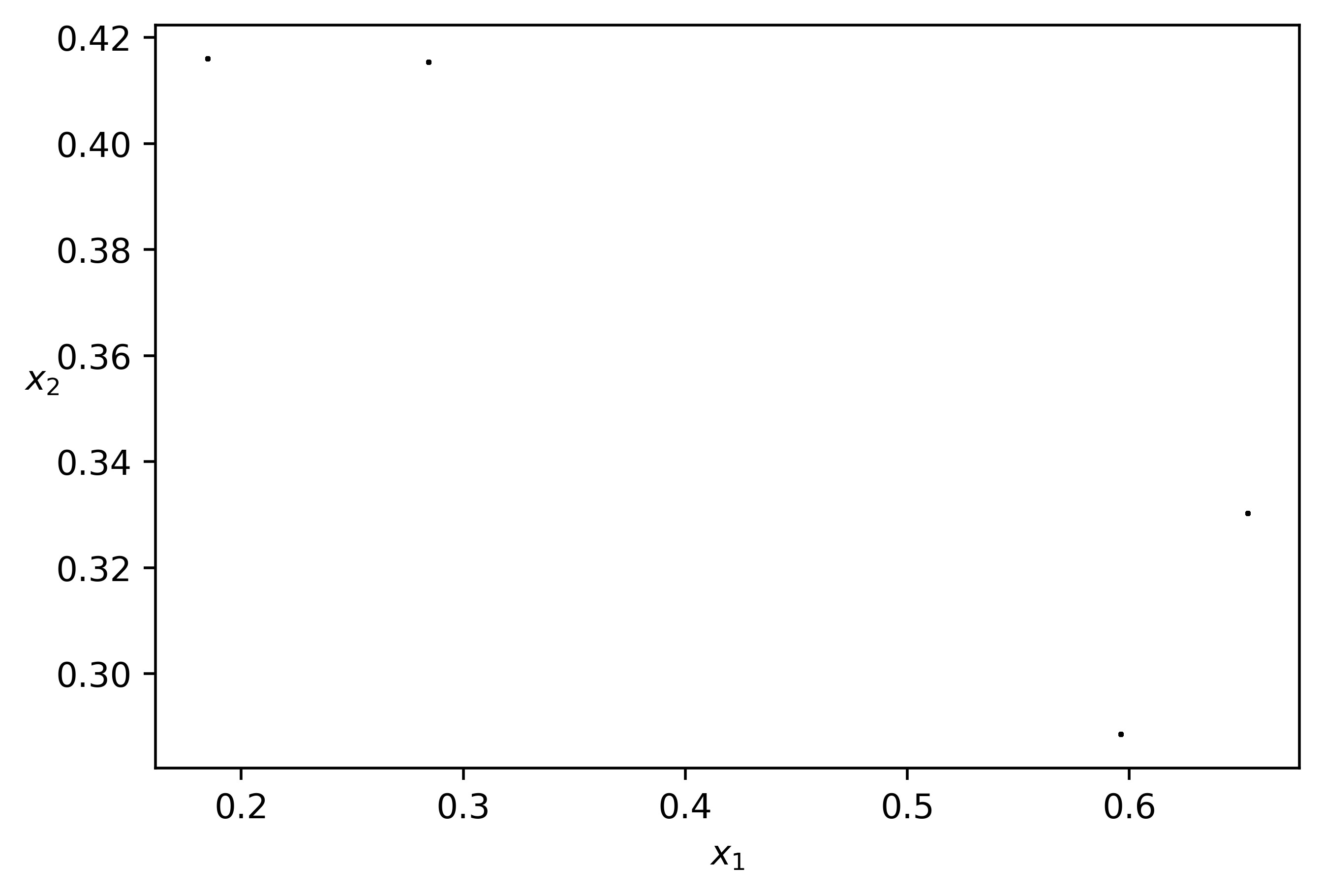}} 
    \subfigure[$\delta =2.368537$.]{\includegraphics[width=0.4\textwidth]{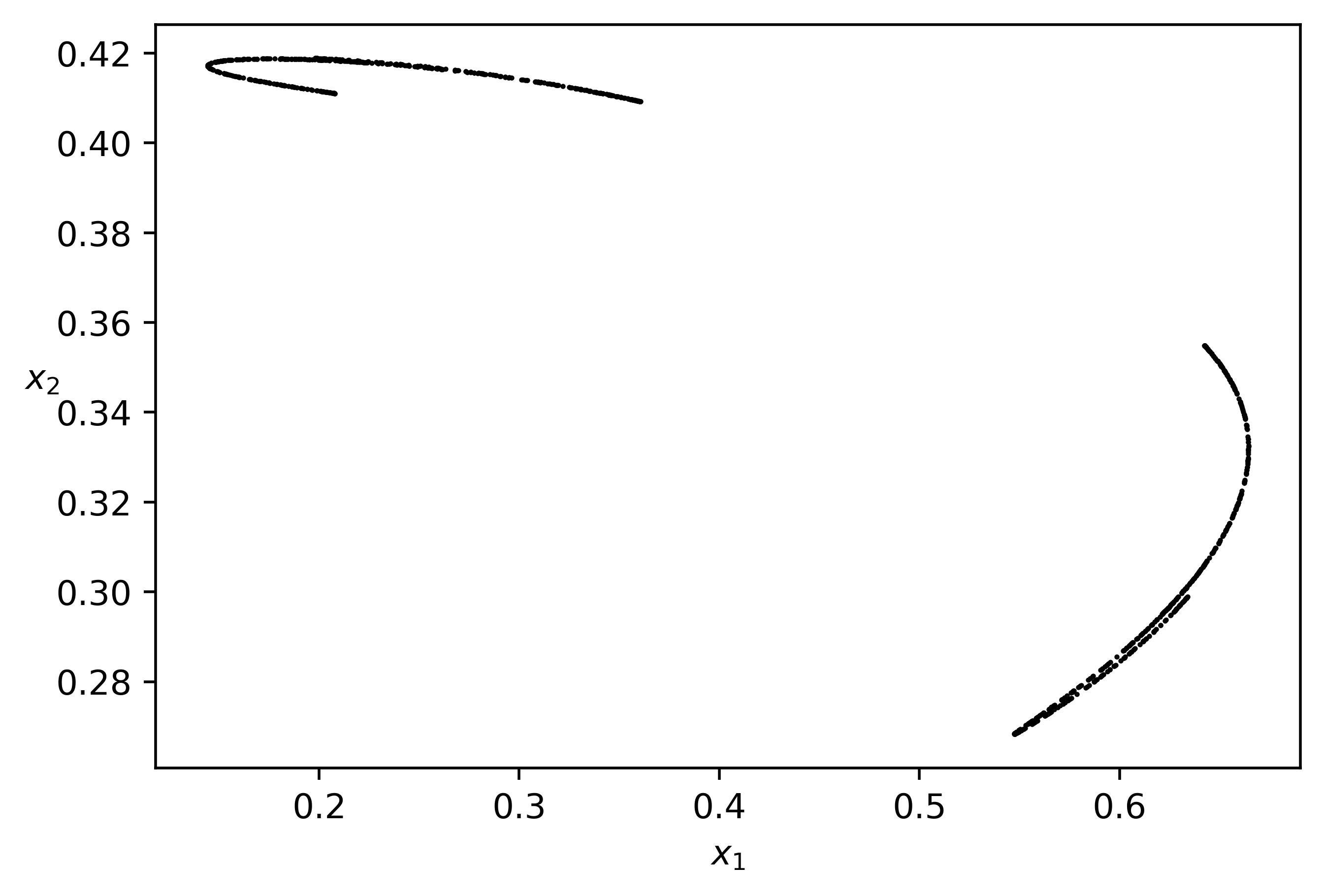}}\\
    \subfigure[$\delta =2.369339$.]{\includegraphics[width=0.4\textwidth]{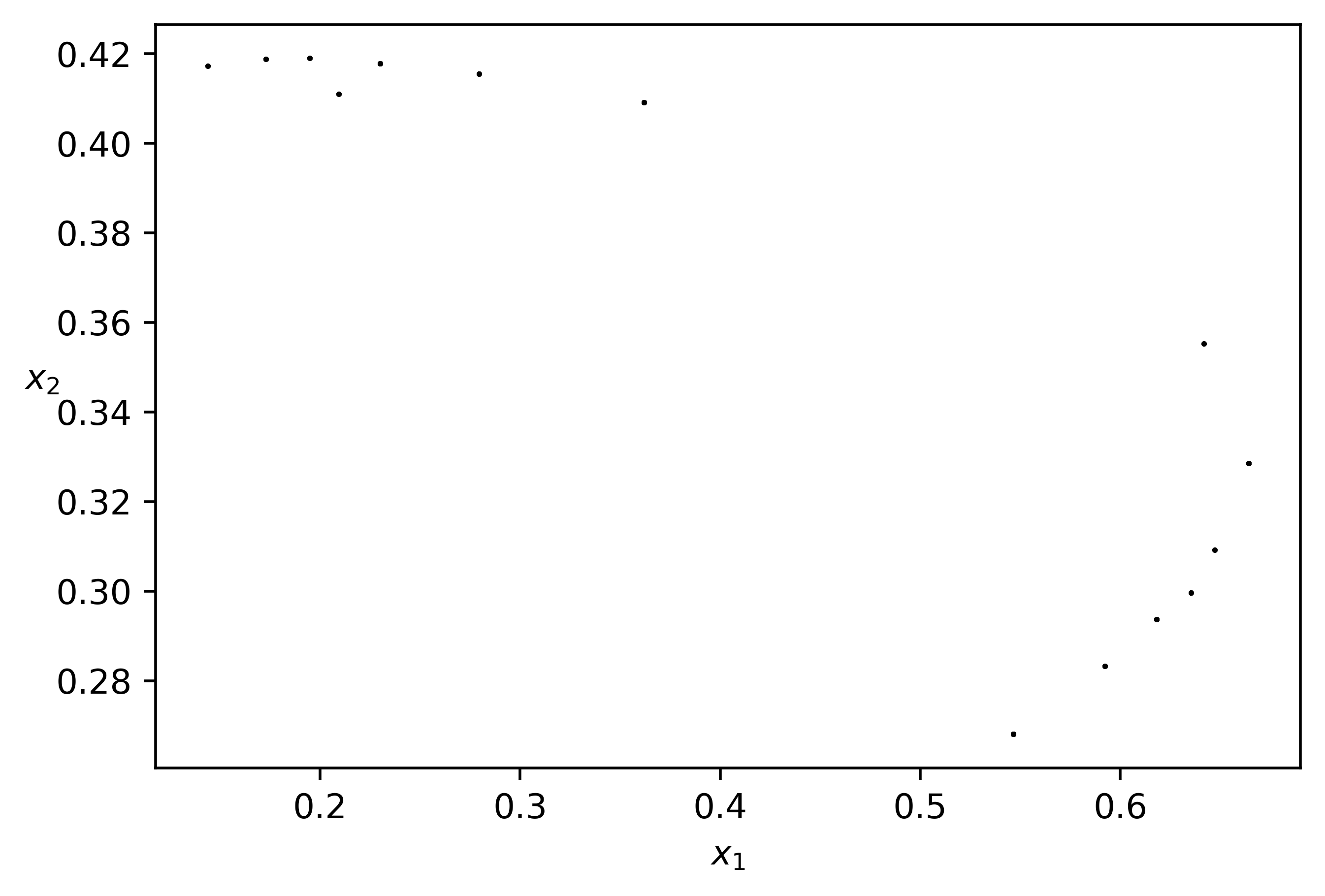}} 
    \subfigure[$\delta =2.479559$.]{\includegraphics[width=0.4\textwidth]{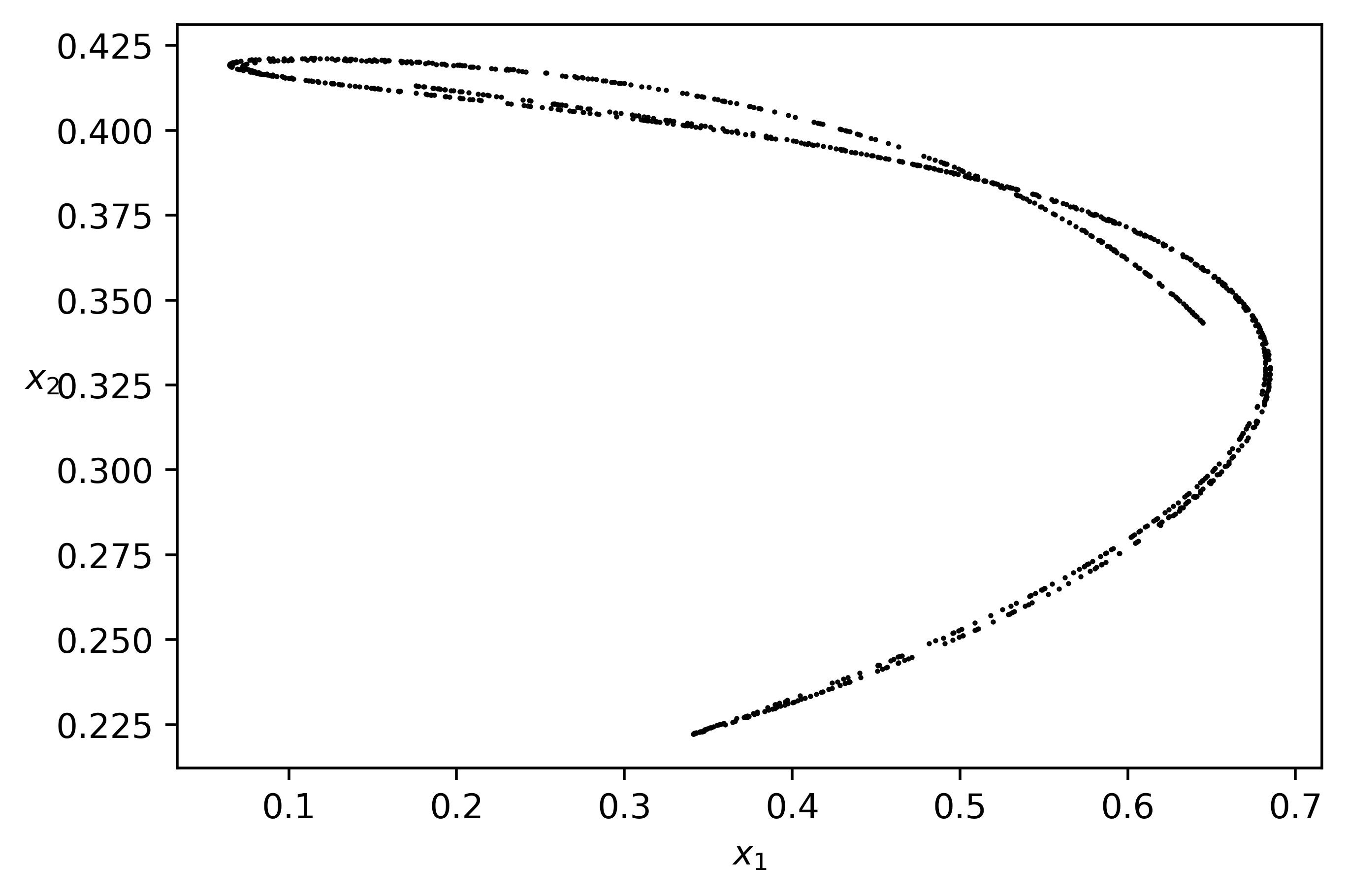}}
  \caption{The phase portraits of Model GL by fixing the parameters $c_1=0.5$, $c_2=1.0$. We choose $(x_1(0),x_2(0))=(0.5,0.5)$ to be the initial state of the iterations.}
    \label{fig:phase-p}
\end{figure}

Concerning Model GL, we plot the phase portraits in Figure \ref{fig:phase-p} by fixing the parameters $c_1=0.5$, $c_2=1.0$ and choosing $(x_1(0),x_2(0))=(0.5,0.5)$ to be the initial state. From Figure \ref{fig:phase-p} (a), one can see that there is a 4-cycle orbit when $\delta =2.300000$. As the value of $\delta$ increases, two pieces of strange attractor emerge when $\delta =2.368537$, which is shown by Figure \ref{fig:phase-p} (b). Afterward, when $\delta =2.369339$, the two pieces of strange attractors transition into a periodic trajectory with order 14 (see Figure \ref{fig:phase-p} (c)). When $\delta =2.479559$, one unique piece of chaotic attractor appears as Figure \ref{fig:phase-p} (d) reports.

\begin{figure}[htbp]
  \centering
	\includegraphics[width=0.9\textwidth]{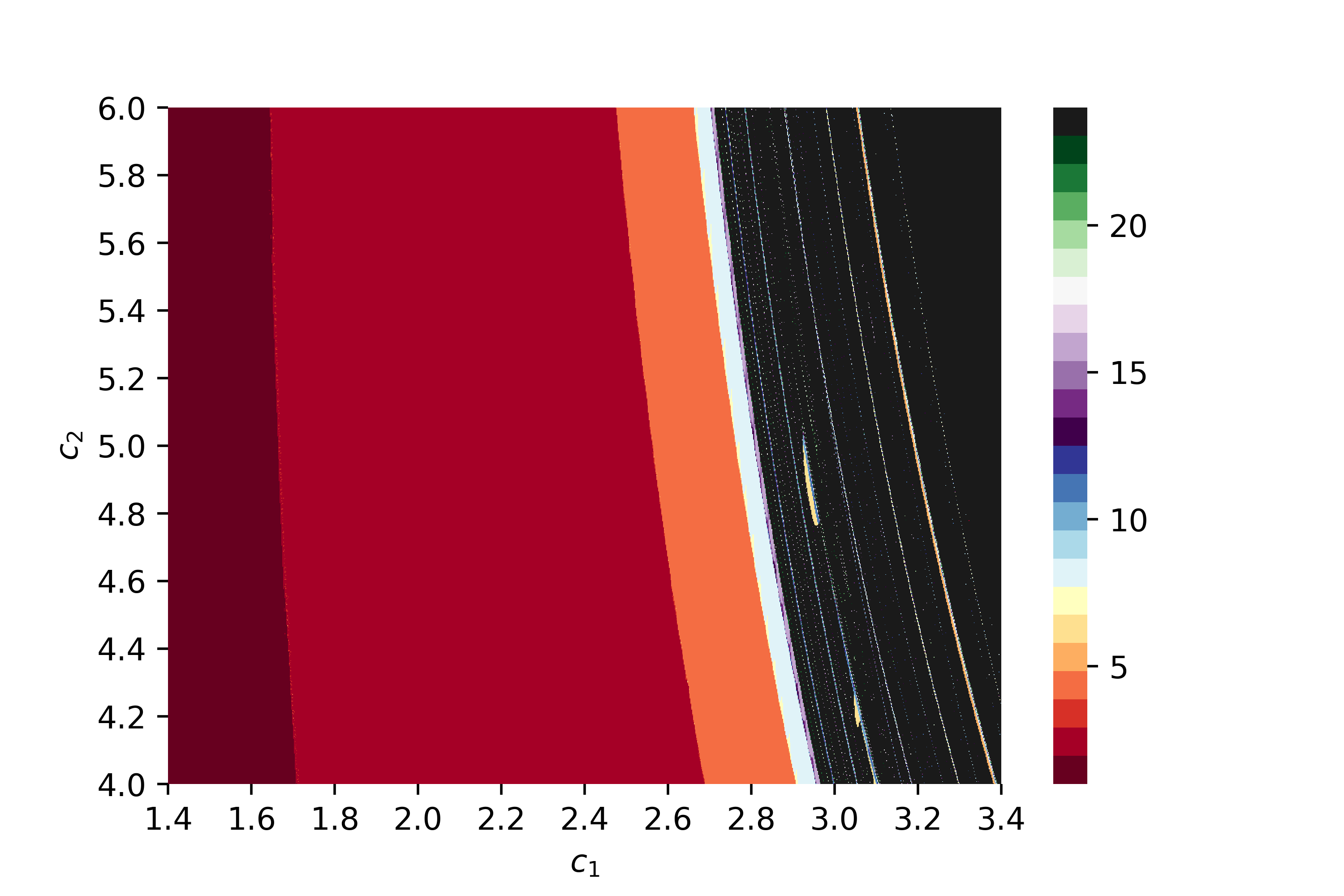}
  \caption{The 2-dimensional bifurcation diagram of Model GL for $(c_1,c_2)\in[1.4,3.4]\times[4.0,6.0]$ by fixing the parameter $\delta=1.0$. We choose $(x_1(0),x_2(0))=(0.2,0.2)$ to be the initial state of the iterations.}
    \label{fig:2d-bif-gl}
\end{figure}

By fixing the parameter $\delta=1.0$ and choosing $(x_1(0),x_2(0))=(0.2,0.2)$ as the initial state, we depict the 2-dimensional bifurcation diagram of Model GL for $(c_1,c_2)\in[1.4,3.4]\times[4.0,6.0]$ in Figure \ref{fig:2d-bif-gl}. Similarly, we use different colors to mark parameter points corresponding to trajectories with different periods. Figure \ref{fig:2d-bif-gl} also shows that Model GL transitions from stability to chaos through period-doubling bifurcations. This confirms the conclusion in Theorem \ref{thm:gl} that the only possible bifurcation for Model GL is the period-doubling bifurcation.

\section{Concluding Remarks}

In this paper, we investigated three Cournot games with two heterogeneous firms, where the second firms are endowed with distinct rationality levels. In our setup, we assumed that the market possesses an isoelastic demand function derived from the Cobb-Douglas preference of consumers. Moreover, the cost functions were supposed to be quadratic. We should mention that both the market demand and firm costs are nonlinear, which allows us to introduce the theory of oligopoly models to more realistic economies, compared to the widely used linear models.

Our study aims at investigating the influence of players' rational degrees on the dynamics of heterogeneous Cournot duopolistic competition. For this purpose, several tools of symbolic computations were employed including, e.g., the triangular decomposition, resultant, and \gai{PCAD method}. Based on symbolic computations, we established rigorous and complete conditions of the local stability and bifurcations for the three models considered in this paper. By comparing the stability regions, we derived that the stability region of GR is the smallest, while that of GB is the largest. It seems that the involvement of the gradient adjustment mechanism leads to the occurrence of these counterintuitive phenomena. However, the reason is still not clear and is left for our future research.

%
%
%

\bibliographystyle{abbrv}
\bibliography{duopoly.bib}

\begin{thebibliography}{10}

\bibitem{Agiza1998E}
H.~Agiza.
\newblock Explicit stability zones for {Cournot} game with 3 and 4 competitors.
\newblock {\em Chaos, Solitons \& Fractals}, 9(12):1955--1966, 1998.

\bibitem{Agliari2016N}
A.~Agliari, A.~Naimzada, and N.~Pecora.
\newblock Nonlinear dynamics of a {Cournot} duopoly game with differentiated
  products.
\newblock {\em Applied Mathematics and Computation}, 281:1--15, 2016.

\bibitem{Ahmed2000O}
E.~Ahmed, H.~Agiza, and S.~Hassan.
\newblock On modifications of {Puu's} dynamical duopoly.
\newblock {\em Chaos, Solitons \& Fractals}, 11(7):1025--1028, 2000.

\bibitem{Askar2016N}
S.~Askar and K.~Alnowibet.
\newblock Nonlinear oligopolistic game with isoelastic demand function:
  Rationality and local monopolistic approximation.
\newblock {\em Chaos, Solitons \& Fractals}, 84:15--22, 2016.

\bibitem{Aubry1999T}
P.~Aubry and M.~Moreno~Maza.
\newblock Triangular sets for solving polynomial systems: a comparative
  implementation of four methods.
\newblock {\em Journal of Symbolic Computation}, 28(1–2):125--154, 1999.

\bibitem{Baiardi2018A}
L.~C. Baiardi and A.~K. Naimzada.
\newblock An oligopoly model with best response and imitation rules.
\newblock {\em Applied Mathematics and Computation}, 336:193--205, 2018.

\bibitem{Bischi2007O}
G.~I. Bischi, A.~K. Naimzada, and L.~Sbragia.
\newblock Oligopoly games with local monopolistic approximation.
\newblock {\em Journal of Economic Behavior \& Organization}, 62(3):371--388,
  2007.

\bibitem{Canovas2018O}
J.~S. C{\'a}novas and M.~Mu{\~n}oz-Guillermo.
\newblock On the dynamics of {Kopel’s} {Cournot} duopoly model.
\newblock {\em Applied Mathematics and Computation}, 330:292--306, 2018.

\bibitem{Cavalli2015Na}
F.~Cavalli and A.~Naimzada.
\newblock Nonlinear dynamics and convergence speed of heterogeneous {Cournot}
  duopolies involving best response mechanisms with different degrees of
  rationality.
\newblock {\em Nonlinear Dynamics}, 81(1):967--979, 2015.

\bibitem{Cavalli2015N}
F.~Cavalli, A.~Naimzada, and F.~Tramontana.
\newblock Nonlinear dynamics and global analysis of a heterogeneous {Cournot}
  duopoly with a local monopolistic approach versus a gradient rule with
  endogenous reactivity.
\newblock {\em Communications in Nonlinear Science and Numerical Simulation},
  23(1-3):245--262, 2015.

\bibitem{Collins1991P}
G.~E. Collins and H.~Hong.
\newblock Partial cylindrical algebraic decomposition for quantifier
  elimination.
\newblock {\em Journal of Symbolic Computation}, 12(3):299--328, 1991.

\bibitem{Cournot1838R}
A.~A. Cournot.
\newblock {\em Recherches sur les Principes Mathématiques de la Théorie des
  Richesses}.
\newblock L. Hachette, Paris, 1838.

\bibitem{Elsadany2012C}
A.~A. Elsadany.
\newblock Competition analysis of a triopoly game with bounded rationality.
\newblock {\em Chaos, Solitons \& Fractals}, 45(11):1343--1348, 2012.

\bibitem{Elsadany2017D}
A.~A. Elsadany.
\newblock Dynamics of a cournot duopoly game with bounded rationality based on
  relative profit maximization.
\newblock {\em Applied Mathematics and Computation}, 294:253--263, 2017.

\bibitem{Fisher1961T}
F.~M. Fisher.
\newblock The stability of the {Cournot} oligopoly solution: The effects of
  speeds of adjustment and increasing marginal costs.
\newblock {\em The Review of Economic Studies}, 28(2):125--135, 1961.

\bibitem{Hommes2013B}
C.~Hommes.
\newblock {\em Behavioral Rationality and Heterogeneous Expectations in Complex
  Economic Systems}.
\newblock Cambridge University Press, 2013.

\bibitem{Huang2019A}
B.~Huang and W.~Niu.
\newblock Analysis of snapback repellers using methods of symbolic computation.
\newblock {\em International Journal of Bifurcation and Chaos}, 29(04):1950054,
  2019.

\bibitem{Jury1976I}
E.~Jury, L.~Stark, and V.~Krishnan.
\newblock Inners and stability of dynamic systems.
\newblock {\em IEEE Transactions on Systems, Man, and Cybernetics},
  (10):724--725, 1976.

\bibitem{Kalkbrener1993A}
M.~Kalkbrener.
\newblock A generalized {Euclidean} algorithm for computing triangular
  representations of algebraic varieties.
\newblock {\em Journal of Symbolic Computation}, 15(2):143--167, 1993.

\bibitem{Kopel1996S}
M.~Kopel.
\newblock Simple and complex adjustment dynamics in {Cournot} duopoly models.
\newblock {\em Chaos, Solitons \& Fractals}, 7(12):2031--2048, 1996.

\bibitem{Li2020N}
B.~Li, Q.~He, and R.~Chen.
\newblock Neimark–{Sacker} bifurcation and the generate cases of {Kopel}
  oligopoly model with different adjustment speed.
\newblock {\em Advances in Difference Equations}, 2020(1):72, 2020.

\bibitem{Li2022C}
B.~Li, H.~Liang, L.~Shi, and Q.~He.
\newblock Complex dynamics of {Kopel} model with nonsymmetric response between
  oligopolists.
\newblock {\em Chaos, Solitons \& Fractals}, 156:111860, 2022.

\bibitem{Li1975P}
T.-Y. Li and J.~A. Yorke.
\newblock Period three implies chaos.
\newblock {\em The American Mathematical Monthly}, 82(10):985--992, 1975.

\bibitem{Li2010D}
X.~Li, C.~Mou, and D.~Wang.
\newblock Decomposing polynomial sets into simple sets over finite fields: The
  zero-dimensional case.
\newblock {\em Computers \& Mathematics with Applications}, 60(11):2983--2997,
  2010.

\bibitem{Li2022A}
X.~Li and L.~Su.
\newblock A heterogeneous duopoly game under an isoelastic demand and
  diseconomies of scale.
\newblock {\em Fractal and Fractional}, 6(8):459, 2022.

\bibitem{Li2014C}
X.~Li and D.~Wang.
\newblock Computing equilibria of semi-algebraic economies using triangular
  decomposition and real solution classification.
\newblock {\em Journal of Mathematical Economics}, 54:48--58, 2014.

\bibitem{Ma2013T}
J.~Ma and X.~Pu.
\newblock The research on {Cournot-Bertrand} duopoly model with heterogeneous
  goods and its complex characteristics.
\newblock {\em Nonlinear Dynamics}, 72(4):895--903, 2013.

\bibitem{Matouk2017N}
A.~E. Matouk, A.~A. Elsadany, and B.~Xin.
\newblock {Neimark-Sacker} bifurcation analysis and complex nonlinear dynamics
  in a heterogeneous quadropoly game with an isoelastic demand function.
\newblock {\em Nonlinear Dynamics}, 89(4):2533--2552, 2017.

\bibitem{Matsumoto2015D}
A.~Matsumoto and F.~Szidarovszky.
\newblock Dynamic monopoly with multiple continuously distributed time delays.
\newblock {\em Mathematics and Computers in Simulation}, 108:99--118, 2015.

\bibitem{Mishra1993A}
B.~Mishra.
\newblock {\em Algorithmic Algebra}.
\newblock Springer-Verlag, New York, 1993.

\bibitem{Naimzada2009C}
A.~Naimzada and F.~Tramontana.
\newblock Controlling chaos through local knowledge.
\newblock {\em Chaos, Solitons \& Fractals}, 42(4):2439--2449, 2009.

\bibitem{Puu1991C}
T.~Puu.
\newblock Chaos in duopoly pricing.
\newblock {\em Chaos, Solitons \& Fractals}, 1(6):573--581, 1991.

\bibitem{Theocharis1960O}
R.~D. Theocharis.
\newblock On the stability of the {Cournot} solution on the oligopoly problem.
\newblock {\em The Review of Economic Studies}, 27(2):133--134, 1960.

\bibitem{Tramontana2010H}
F.~Tramontana.
\newblock Heterogeneous duopoly with isoelastic demand function.
\newblock {\em Economic Modelling}, 27(1):350--357, 2010.

\bibitem{Wang2000C}
D.~Wang.
\newblock Computing triangular systems and regular systems.
\newblock {\em Journal of Symbolic Computation}, 30(2):221--236, 2000.

\bibitem{Wu2010C}
W.~Wu, Z.~Chen, and W.~Ip.
\newblock Complex nonlinear dynamics and controlling chaos in a {Cournot}
  duopoly economic model.
\newblock {\em Nonlinear Analysis: Real World Applications}, 11(5):4363--4377,
  2010.

\bibitem{Wu1986B}
W.-T. Wu.
\newblock Basic principles of mechanical theorem proving in elementary
  geometries.
\newblock {\em Journal of Automated Reasoning}, 2(3):221--252, 1986.

\end{thebibliography}

%
%


%
%
%
%
%
%
%
%
%
%
%
%
%
%
%
%

\end{CJK}
\end{document}